\documentclass[useAMS,usenatbib]{mn2e}
\usepackage{times}
\usepackage{verbatim}
\usepackage{amsmath}
\usepackage{graphicx}
\usepackage{wrapfig} 
\usepackage{caption}
\usepackage{hyperref} 
\usepackage[3D]{movie15}

\usepackage{color}
\usepackage{aas_macros}
\usepackage{graphicx}
\usepackage{epsf}
\usepackage{graphics}
\usepackage{rotating}
\usepackage{amssymb}   
\usepackage{setspace}  
\usepackage{lscape}

\newcommand{\be}{\begin{equation}}
\newcommand{\ee}{\end{equation}}

\newcommand{\kms}{\,km\,s$^{-1}$}
\newcommand{\mpc}{$h^{-1}$Mpc}

\defcitealias{lynden-bell88}{LB88}

\title{The 6dF Galaxy Survey: Peculiar Velocity Field and Cosmography}

\author[Springob et~al.]
{Christopher M. Springob$^{1,2,3}$, Christina Magoulas$^{4,5,3}$,
Matthew Colless$^{6}$, Jeremy Mould$^{2,7}$, 
\newauthor
Pirin Erdo{\u g}du$^{8}$, D. Heath Jones$^{9}$, John R. Lucey$^{10}$, 
Lachlan Campbell$^{3}$, Christopher J. Fluke$^{7}$\\
$^1$International Centre for Radio Astronomy Research, The University of
Western Australia, Crawley, WA 6009 Australia\\ 
$^2$ARC Centre of Excellence for All-sky Astrophysics (CAASTRO)\\
$^3$Australian Astronomical Observatory, PO Box 915, North Ryde, NSW
1670, Australia\\ 
$^4$School of Physics, University of Melbourne, Parkville, VIC 3010,
Australia\\ 
$^5$Department of Astronomy, University of Cape Town, Private Bag X3, Rondebosch, 7701, South Africa\\
$^6$Research School of Astronomy and Astrophysics, The Australian
National University, Canberra, ACT 2611, Australia\\ 
$^7$Centre for Astrophysics and Supercomputing, Swinburne University,
Hawthorn, VIC 3122, Australia\\ 
$^8$Australian College of Kuwait, PO Box 1411, Safat 13015, Kuwait\\ 
$^9$School of Physics, Monash University, Clayton, VIC 3800, Australia\\ 
$^{10}$Department of Physics, University of Durham, Durham DH1 3LE, UK\\ 
}

\begin{document}

\maketitle

\begin{abstract}
We derive peculiar velocities for the 6dF Galaxy Survey (6dFGS) and describe the velocity field of the nearby ($z<0.055$) southern hemisphere.  The survey comprises 8885 galaxies for which we have previously reported Fundamental Plane data.  We obtain peculiar velocity probability distributions for the redshift space positions of each of these galaxies using a Bayesian approach.  Accounting for selection bias, we find that the logarithmic distance uncertainty is 0.11 dex, corresponding to $26\%$ in linear distance.  We use adaptive kernel smoothing to map the observed 6dFGS velocity field out to $cz \sim 16,000$ \kms, and compare this to the predicted velocity fields from the PSCz Survey and the 2MASS Redshift Survey.  We find a better fit to the PSCz prediction, although the reduced $\chi^2$ for the whole sample is approximately unity for both comparisons.  This means that, within the observational uncertainties due to redshift independent distance errors, observed galaxy velocities and those predicted by the linear approximation from the density field agree.  However, we find peculiar velocities that are systematically more positive than model predictions in the direction of the Shapley and Vela superclusters, and systematically more negative than model predictions in the direction of the Pisces-Cetus Supercluster, suggesting contributions from volumes not covered by the models.
\end{abstract}

\section{Introduction}

{\it [Note: This is the 2D version of this paper.  For readers using Adobe Reader 8.0 or higher, we recommend viewing the 3D version, which includes 3D interactive copies of Figures 11 and 12.  Figures 11 and 12 can be downloaded as ancillary files from astro-ph, while the complete paper with embedded 3D plots should also be available from http://www.6dfgs.net/vfield/veldata.pdf .]}

The velocity field of galaxies exhibits deviations from Hubble flow
induced by inhomogeneities in the large scale distribution of matter. By
studying the galaxy peculiar velocity field, we can explore the large
scale distribution of matter in the local universe and so test
cosmological models and measure cosmological parameters.

The measurement of galaxy peculiar velocities involves evaluating both the
redshifts and distances of galaxies, and computing the residual
component of the velocity that is not accounted for by Hubble flow. 
The peculiar velocity is defined as 
\be v_{pec} \equiv cz_{pec} \ee 
where the peculiar redshift $z_{pec}$ is related to the observed redshift
$z_{obs}$ and the redshift due to the Hubble flow $z_{H}$ through
\be (1+z_{obs}) = (1+z_{H})(1+z_{pec}) ~.\ee
(See \citealt{harrison74}.)  At low redshifts, the peculiar velocity approximates to
\be v_{pec} \approx cz_{obs} - cz_{H} \approx cz_{obs} - H_0 D \ee
where $H_0$ is the Hubble constant and $D$ is the galaxy's comoving distance.  Throughout this paper, we use the exact relation (Equation 2) rather than this approximation.

The measurement of the peculiar velocities thus depends on the use of
redshift-independent distance indicators. Many distance indicators have
been used over the years (see \citealt{jacoby92} for an overview of several of these indicators), but the two that have yielded the largest
number of distance measurements are the Tully-Fisher relation (TF; \citealt{tully77}) and the Fundamental Plane relation (FP; \citealt{dressler87a}, \citealt{djorgovski87x}). The former is a scaling relation for
late-type galaxies that expresses the luminosity as a
power law function of rotation velocity. The latter is a scaling
relation for galaxy spheroids (including spiral bulges) that expresses the
effective radius as a power-law product of effective
surface brightness and central velocity dispersion.

The earliest wide-angle peculiar velocity surveys included several
hundred galaxies. Many of these surveys were combined to
create the Mark III catalog (\citealt{willick95}, \citealt{willick96}, \citealt{willick97}). The
earliest FP peculiar velocity surveys to include more than 1000 galaxies
were ENEAR (\citealt{dacosta00a}, \citealt{bernardi02a}), EFAR (\citealt{colless01a}, \citealt{saglia01}), and the Streaming Motions of Abell
Clusters survey \citep{hudson01}. The earliest TF
peculiar velocity surveys of comparable size were a set of
overlapping surveys conducted by Giovanelli, Haynes, and collaborators
(e.g., \citealt{giovanelli94}, \citealt{giovanelli95}, \citealt{giovanelli97}, \citealt{haynes99a}, \citealt{haynes99b}).

The largest TF survey used for peculiar velocity studies to date (and
the largest single peculiar velocity survey published until now) is the
SFI++ survey (\citealt{masters06x}, \citealt{springob07}), which included
TF data for $\sim 5000$ galaxies (much of which came from the earlier SFI, SCI,
and SC2 surveys). SFI++ has been included,
along with other surveys using additional techniques, into yet larger catalogs of peculiar velocities, such as the 
the COMPOSITE sample \citep{watkins09} and the Extragalactic Distance Database \citep{tully09}.

Peculiar velocity surveys have long been used for cosmological
investigations. In addition they have also been used to
study the cosmography of the local universe. Because the existing sample
of galaxy peculiar velocities remains sparse, the most detailed
cosmographic description of the velocity field has been confined to the
nearest distances. Most significantly, the Cosmic Flows survey (\citealt{courtois11b}, \citealt{courtois11a}) has been used to
investigate the cosmography of the velocity field within 3000\kms\
\citep{courtois12a}.  This has now been extended with the followup Cosmic Flows 2 survey (\citealt{tully13}).  Cosmographic descriptions of the velocity field
at more distant redshifts have been made, though the sampling of the
larger volumes is sparse (e.g., \citealt{hudson04}).  Perhaps the most extensive examination of the cosmography of the local universe to somewhat higher redshifts was done by \citet{theureau07}, who looked at the velocity field out to 8000\kms\ using the Kinematics of the Local Universe sample (\citealt{theureau05}, and references therein).

One focus of study has been the comparison of peculiar velocity field
models derived from redshift surveys to the observed peculiar velocity
field. Early comparisons
involved models based on the expected infall around one or more large
attractors (e.g., \citealt{lynden-bell88}, \citealt{han90}, \citealt{mould00}). The subsequent advent of large all-sky redshift surveys allowed various
authors to reconstruct the predicted velocity field from the redshift
space distribution of galaxies, treating every individual galaxy as an
attractor. That is, the velocity field was reconstructed under the
assumption that the galaxy density field traced the underlying matter density field,
assuming a linear bias parameter $b=\delta_{g}/\delta_{m}$, where
$\delta_g$ and $\delta_m$ represent the relative overdensity in the
galaxy and mass distributions respectively.

Early attempts to compare the observed peculiar velocity
field to the field predicted by large all-sky redshift surveys include \citet{kaiser91}, \citet{shaya92}, \citet{hudson94}, and \citet{davis96}.  Subsequent studies exploited the deeper density/velocity field reconstruction of the IRAS Point Source Catalogue Redshift Survey (PSCz, \citealt{saunders00}) by \citet{branchini99}, e.g.
\citet{nusser01}, \citet{branchini01}, \citet{hudson04},
\citet{radburn04}, \citet{ma12}, and \citet{turnbull12}.
The density/velocity field reconstructions have also been derived using galaxy samples selected from the 2MASS XSC catalog \citep{jarrett00x}, e.g. \citet{pike05}, \citet{erdogdu06}, \citet{lavaux10}, \citet{davis11a}. Recently Erdo{\u g}du et al. (2014, submitted),
using the deeper $K_s = 11.75$ limited version of the 2MASS Redshift Survey (2MRS, \citealt{huchra12}), have derived an updated reconstruction of the 2MASS density/velocity field.

The various density and velocity field reconstructions are able to
recover all of the familiar features of large scale structures apparent
in redshift surveys, though there are some disagreements at smaller scales. Additionally, the question of whether the
velocity field reconstructions can replicate the full CMB dipole remains
unresolved, and the degree of agreement between the dipole of the
observed velocity field and both $\Lambda$CDM predictions and the
reconstructed velocity fields from redshift surveys remains in dispute
(e.g., \citealt{feldman10}; \citealt{nusser11}).

Deeper redshift and peculiar velocity surveys could help
to resolve these issues and give us a better understanding of the
cosmography of the local universe. Most of the deeper surveys to date
include either a very small number of objects or heterogeneous selection
criteria. Real gains can be made from a deep peculiar velocity survey with a large number of uniformly selected objects. In this
paper, we present the results from just such a survey: the 6-degree
Field Galaxy Survey (6dFGS).

6dFGS is a combined redshift and peculiar velocity survey of galaxies
covering the entire southern sky at $|b|>10^{\circ}$ (\citealt{jones04}, \citealt{jones05x}, \citealt{jones09x}). The redshift survey includes more than 125,000 galaxies and
the peculiar velocity subsample (hereafter 6dFGSv) includes $\sim$10,000
galaxies, extending in redshift to $cz \approx 16,000$\kms. This is the
largest peculiar velocity sample from a single survey to date.

The peculiar velocities are derived from FP data for these galaxies.
The spectroscopic observations were made with the UK Schmidt Telescope, and photometric observations come from the
Two Micron All-Sky Survey (2MASS) Extended Source Catalog \citep{jarrett00x}. When plotted in the 3-dimensional parameter space with axes
$r=log(R_e)$, $s=log(\sigma_0)$, and $i=log(I_e)$, where $R_e$,
$\sigma_0$, and $I_e$ represent effective radius, central velocity
dispersion, and effective surface brightness respectively, the galaxies
lie along a plane that can be expressed in the form
\be r=as+bi+c \ee
where $a$, $b$ and $c$ are observationally derived constants. Because
$r$ is a distance-dependent quantity while both $s$ and $i$ are
essentially distance-independent, the FP can be used as a distance
indicator, with the galaxy's FP offset along the $r$-direction
providing a measure of its peculiar velocity.

The final data release for 6dFGS redshifts was presented by \citet{jones09x}. The data release for the FP parameters was \citet{campbell14}. The fitting of the FP is described by
\citet{magoulas12x}, while the stellar population trends in FP space were examined by \citet{springob12}.

In this paper, we present the method for deriving the peculiar
velocities for the 6dFGSv galaxies, and we provide an overview of the
peculiar velocity cosmography, which will inform the cosmological
analyses that we will undertake in future papers.  These papers include a measurement of the growth rate of structure \citep{johnson14} and measurements of the bulk flow, using different methods (Magoulas et al., in prep., Scrimgeour et al., submitted).

This paper is arranged as follows. In Section 2 we describe both the
6dFGSv dataset and the 2MRS and PSCz predicted velocity fields to which we will
compare our results. In Section 3 we describe the fitting of the FP and
in Section 4 we describe the derivation of the peculiar velocities. In
Section 5 we discuss our adaptive kernel smoothing, and the resulting
6dFGSv cosmography. Our results are summarized in Section 6.

\section{Data}

\subsection{6dFGSv Fundamental Plane data}

The details of the sample selection and data reduction are presented in
\citet{magoulas12x} and \citet{campbell14}. In brief, the
6dFGSv includes all 6dFGS early-type galaxies with spectral
signal-to-noise ratios greater than 5, heliocentric redshift
$z_{helio}<0.055$, velocity dispersion greater than 112\kms, and J-band
total magnitude brighter than $m_J = 13.65$. The galaxies were
identified as `early-type' by matching the observed spectrum, via
cross-correlation, to template galaxy spectra. They include both ellipticals and spiral bulges (in cases where the bulge
fills the 6dF fibre). Each galaxy image was subsequently examined by
eye, and galaxies were removed from the sample in cases where the morphology was
peculiar, the galaxy had an obvious dust lane, or the fibre aperture was
contaminated by the galaxy's disk (if present), or by a star or another galaxy.

\begin{figure*}
\begin{minipage}{175mm}
\includegraphics[width=1.0\textwidth]{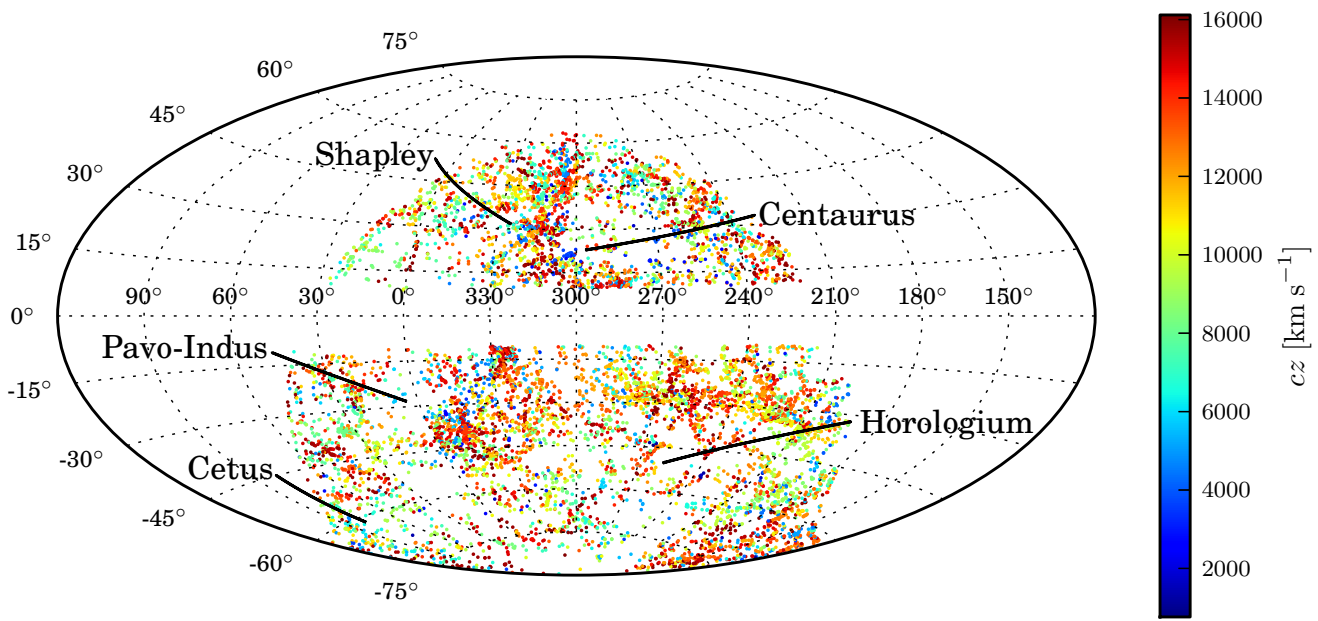}
\caption{Distribution of 6dFGSv galaxies in Galactic latitude ({\it l})
  and longitude ({\it b}), shown in an equal-area Aitoff projection.
  Individual galaxies are color-coded by their redshift. The 6dFGSv
  galaxies fill the southern hemisphere apart from $\pm$10$^\circ$ about
  the Galactic plane. Some of the large scale structures in the 6dFGSv
  volume are also indicated.
\label{FIG1}}
\end{minipage}
\end{figure*}

We have also removed from the sample several hundred galaxies within the
heliocentric redshift limit of $z_{helio}=0.055$ that nonetheless have
recessional velocities greater than 16,120\kms\ in the Cosmic Microwave
Background (CMB) reference frame. We do this because our peculiar
velocity analysis is done in the CMB frame, and we wish the survey to
cover a symmetric volume in that frame.
Since the initial survey redshift limit was made in the heliocentric
frame, we must limit the sample to 16,120\kms\ in the CMB frame in order
to have a uniform redshift limit across the sky.  The final sample has 8885 galaxies.

Velocity dispersions were measured from the 6dFGS spectra, using the
Fourier cross-correlation method of \citet{tonry79}. The method
involves convolving the galaxy spectrum with a range of high
signal-to-noise stellar templates, which were also observed with the 6dF
spectrograph.  From that cross-correlated spectrum, we measure the velocity dispersion.  As we demonstrate in \citet{campbell14}, in cases where a galaxy's velocity dispersion has been previously published in the literature, our measurements are in good agreement with the literature values.

The apparent magnitudes were taken from the 2MASS Extended Source
Catalog \citep{jarrett00x}. We have derived the angular radii and
surface brightnesses from the 2MASS images in J-, H-, and K-bands for
each of the galaxies in the sample, taking the total magnitudes from the
2MASS catalog, and then measuring the location of the isophote that
corresponds to the half light radius. Surface brightness as defined here
is then taken to be the average surface brightness interior to the half
light radius. We use the J-band values here, as they offer the smallest
photometric errors.  Again, as shown in \citet{campbell14}, in cases where previously published photometric parameters are available, our measurements are in good agreement.

For the purpose of fitting the Fundamental Plane, the angular radii have been
converted to physical radii using the angular diameter distance corresponding to the observed redshift in the CMB frame.  2666 of the galaxies are in groups or clusters,
as defined by the grouping algorithm outlined by \citet{magoulas12x}.
For these galaxies, we use the redshift distance of the group or
cluster, where the group redshift is defined as the median redshift
for all galaxies in the group.

Several changes have been made to the 6dFGS catalog since the earliest 6dFGS FP papers, \citet{magoulas12x} and \citet{springob12}, were published.  First, the velocity dispersion errors are now derived using a bootstrap technique.  Second, the Galactic extinction corrections are applied using the values given by \citet{schlafly11} rather than \citet{schlegel98}.  Third, $\sim 100$ galaxies with photometric problems (e.g., either 2MASS processing removed a substantial part of the target galaxy or the presence of a strong core asymmetry indicated multiple structures) have been removed from the sample.  These revisions are discussed in greater detail by \citet{campbell14}.  Following these changes, the Fundamental Plane has been re-fit, and the revised FP is discussed in Section 3.

The 6dFGSv sky distribution is shown in Figure~1.
Each point represents a 6dFGSv galaxy, color-coded by redshift. As seen here, 6dFGSv fills the Southern
Hemisphere outside the Zone of Avoidance. Figure~2 shows the
redshift distribution for 6dFGSv, in the CMB frame (which we use
throughout the rest of this paper). As the figure makes clear, the
number of objects per unit redshift increases up to the redshift limit
of the sample. The mean redshift of the sample is 11,175\kms.  Note that a complete, volume-limited sample would have a quadratic increase in the number of objects per redshift bin, and thus a mean redshift of 12,090\kms , or 0.75 times the limiting redshift.


\subsection{Reconstructed velocity fields}

We wish to
compare our observed velocity field to reconstructed velocity field models derived from the
redshift space distribution of galaxies, under the assumption that the
matter distribution traces the galaxy distribution.  We present here two different velocity field reconstructions, one derived from the 2MASS Redshift Survey (Erdo{\u g}du et al., submitted), and one derived from the PSCz survey \citep{branchini99}.  In a future paper, we will also compare the observed velocity field to other reconstructions, including the 2M++ reconstruction \citep{lavaux11}.  This follows Carrick et al. (submitted), who have made such a comparison between 2M++ and SFI++.


\subsubsection{2MRS reconstructed velocity field}

At present, one of the largest and most complete reconstructed velocity
fields is derived from galaxies in the 2MASS Redshift
Survey (2MRS). In the final data release \citep{huchra12}, the 2MRS
consists of redshifts for 44,699 galaxies with a magnitude
limit of $K_s = 11.75$ (with a significant fraction of the southern hemisphere redshifts coming
from 6dFGS).  The zone of avoidance for the sample varies with Galactic longitude, but lies at roughly $|b| \sim 5-8^{\circ}$, and the sample covers $91\%$ of the sky.  We thus make use of the 2MRS
reconstructed density and velocity fields of Erdo{\u g}du et~al.\ (2014,
submitted; updated from \citealt{erdogdu06}) which uses the 2MRS
redshift sample to recover the linear theory predictions for density and
velocity.

The
method of reconstruction is outlined in \citet{erdogdu06}, where it
was applied to a smaller 2MRS sample of 20,860 galaxies with a brighter
magnitude limit of $K_s = 11.25$ and a median redshift of 6000\kms. The
method closely follows that of \citet{fisher95b} and relies on the
assumption that the matter distribution traces the galaxy distribution
in 2MRS, with a bias parameter $\beta = \Omega_m^{0.55} / b$ that is assumed to take the value 0.4 for the 2MRS sample. The
density field in redshift space is decomposed into spherical harmonics
and Bessel functions (or Fourier-Bessel functions) and smoothed using a
Wiener filter. The velocity field is derived from the Wiener-filtered
density field by relating the harmonics of
the gravity field to those of the density field (in linear theory). The
reconstruction gives velocity vectors on a grid in supergalactic
cartesian coordinates with gridpoints spaced by 8 \mpc \ and extending
to a distance of 200 \mpc \ from the origin in each direction.  ($h$ is the Hubble constant in
units of 100\kms\,Mpc$^{-1}$.)

\begin{figure}
\includegraphics[width=0.45\textwidth]{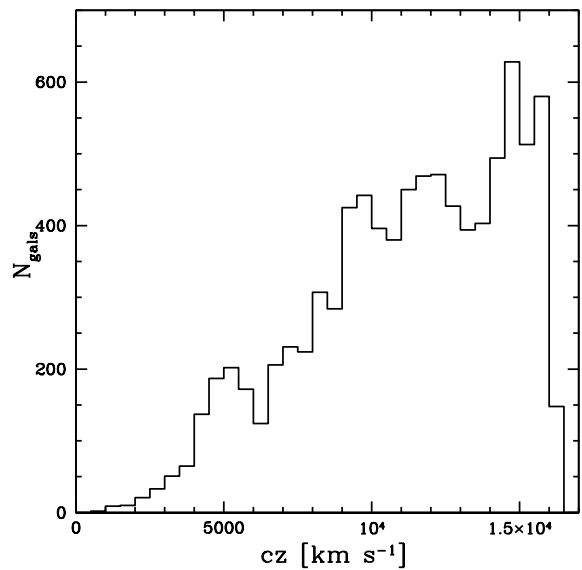}
\caption{Redshift distribution of galaxies in 6dFGSv in the CMB
  reference frame.  The bin width is 500 \kms.
\label{FIG2}}
\end{figure}

\citet{erdogdu06} explore the issue of setting boundary conditions in the density and velocity field reconstruction.  One must make some assumptions about the calibration of the reconstructed density field.  The density/velocity field reconstruction used in this paper defines the logarithmic derivative of the gravitational potential to be continuous along the surface of the sphere of radius 200 \mpc .  This is the ``zero potential'' boundary condition, as described in the aforementioned papers.  For a perfectly smooth, homogeneous universe, one would then expect that both the mean overdensity and mean peculiar velocity within the spherical survey volume would also be zero.  However, because of the particular geometry of local large scale structure, both of these quantities deviate slightly from zero.  The mean overdensity within 200 \mpc \ is found to be $\delta=+0.09$ (with an rms scatter of 1.21), and the mean line-of-sight velocity is found to be +66 \kms (with an rms scatter of 266 \kms ).  (Here, $\delta$ represents the local matter density contrast.)  In contrast to what one might naively expect, the slightly positive mean value of $\delta$ induces a positive value to the mean line-of-sight peculiar velocity.  This occurs because many of the largest structures lie at the periphery of the survey volume.

For comparison with our observed velocities, we convert
the 2MRS velocity grid from real space to redshift space. Each
real-space gridpoint is assigned to its corresponding position in
redshift space, and resampled onto a regularly spaced
grid in redshift space. The points on the redshift-space grid are
4 \mpc \ apart, and we have linearly interpolated the nearest points
from the old grid onto the new grid to get the redshift space
velocities.

Because the real-space velocity field is Weiner-filtered onto a coarsely
sampled grid with 8 \mpc \ spacing, there are no apparent triple-valued
regions in the field. That is, there are no lines of sight along which
the conversion from real space to redshift space becomes confused
because a single redshift corresponds to three different distances, as
can happen in the vicinity of a large overdensity. For triple-valued
regions to appear in a grid with 8 \mpc \ spacing one would need
velocity gradients as large as 800\kms\ between adjacent points in the
grid, and this does not occur anywhere in the velocity field.  While the actual velocity field will presumably include such triple-valued regions around rich clusters, they have been smoothed out in this model.


\subsubsection{PSCz reconstructed velocity field}

An alternative reconstruction of the density and velocity fields is offered by \citet{branchini99}, who make use of the IRAS Point Source Catalog Redshift Survey (PSCz, \citealt{saunders00}).  PSCz includes 15,500 galaxies, with $60 \mu$m flux $f_{60}>0.6$.  The survey covers $84\%$ of the sky, with most of the missing sky area lying at low Galactic latitudes.  (See \citealt{branchini99} Figure 1.)  While the number of galaxies is far fewer than in 2MRS, \citet{erdogdu06} shows that the redshift histogram drops off far more slowly for PSCz than for 2MRS, so that the discrepancy in the number of objects is not as great at distances of $\sim 100-150$ \mpc, where most of our 6dFGSv galaxies lie.

The density and velocity fields were reconstructed from PSCz by spherical harmonic expansion, based on a method proposed by \citet{nusser94}.  The method uses the fact that, in linear theory, the velocity field in redshift space is irrotational, and so may be derived from a velocity potential.  The potential is expanded in spherical harmonics, and the values of the spherical harmonic coefficients are then derived, again assuming a mapping between the PSCz galaxy redshift distribution and the matter distribution, with a bias parameter $\beta = 0.5$.  The reconstruction gives velocity vectors on a supergalactic cartesian grid with spacing 2.8 \mpc, extending to a distance of $180$ \mpc \ from the origin in each direction.  The mean overdensity within the survey volume is $\delta = -0.11$ (with an rms scatter of 1.11), with a mean line-of-sight velocity of +79 \kms (with an rms scatter of 156 \kms ).

We convert the PSCz velocity grid from real space to redshift space, using the same procedure used to construct the 2MRS velocity field.  However, in this instance, the velocity grid uses the same 2.8 \mpc \ grid spacing that is used in the real space field.  While the original grid spacing is finer in the PSCz reconstruction than in the 2MRS reconstruction, \citet{branchini99} minimizes the problem of triple-valued regions by collapsing galaxies within clusters, and applying a method devised by \citet{yahil91} to determine the locations of galaxies along those lines of sight.

\subsubsection{Comparing the model density fields}

The two model density fields are shown in Figure 3.  In the left column, along four different slices of SGZ, we show the reconstructed 2MRS density field.  (SGX, SGY, and SGZ are the three cardinal directions of the supergalactic cartesian coordinate system, with SGZ=0 representing the supergalactic plane.)  The field has been smoothed using a 3D Gaussian kernel, which we explain in greater detail in Section 5, when we apply this smoothing algorithm to the velocity field.  In the central column, we show the smoothed PSCz field, and in the right column, we show the ratio of the 2MRS and PSCz densities.  We only show gridpoints with distances out to 161.2 \mpc, the distance corresponding to the 6dFGSv limiting redshift of 16,120 \kms.

\begin{figure*} \centering 
\begin{minipage}{175mm}
\includegraphics[width=1.0\textwidth]{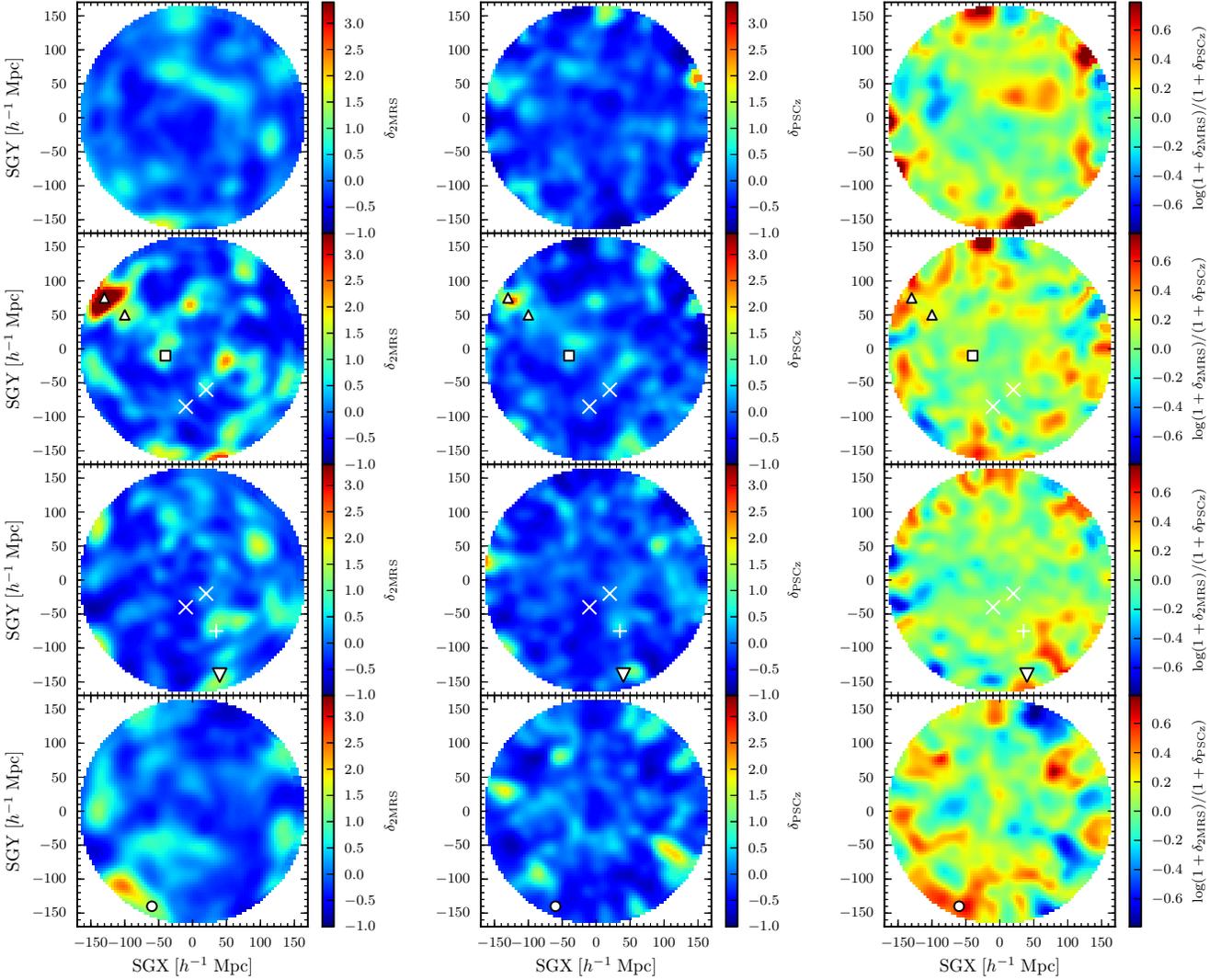} 
\caption{ Gaussian smoothed versions of the 2MRS matter density field
  (left), the PSCz matter density field (centre) and the logarithm of the ratio between the 2MRS and PSCz densities (right) in four slices parallel to the supergalactic plane, covering the ranges (from top to bottom)
  SGZ$>$+20\,$h^{-1}$\,Mpc, $-$20$<$SGZ$<$+20\,$h^{-1}$\,Mpc,
  $-$70$<$SGZ$<$$-$20\,$h^{-1}$\,Mpc and SGZ$<$$-$70\,$h^{-1}$\,Mpc.  Note that $\delta$ is the density {\it contrast}, while $1+\delta$ is the density in units of the mean density of the universe.  The
  slices are arranged so that there are roughly equal numbers of 6dFGSv
  galaxies in each.  Major
  large-scale structures are labelled: the Cetus Supercluster ($\bigtriangledown$), the Eridanus Cluster ($+$), the
  Horologium-Reticulum Supercluster ($\bigcirc$), the Hydra-Centaurus
  Supercluster ($\square$), the four most overdense regions of the
  Sculptor Wall ($\times$), and the two main overdensities of the Shapley
  Supercluster ($\triangle$).  Only gridpoints for which the distance to the origin is less than 161.2 \mpc \ are displayed, so that the limiting distance shown here matches the limiting redshift of 6dFGSv.
\label{FIG3}} 
\end{minipage} 
\end{figure*}

In this figure, we also plot the  positions of several individual Southern Hemisphere superclusters, as identified in the 6dFGS redshift survey.  Based on the features identified by \citet{jones09x}, which itself relies on superclusters identified by \citet{fairall98} and \citet{fairall06}, among others, we highlight the positions of the Cetus Supercluster, Eridanus Cluster, Sculptor Wall, Hydra-Centaurus Supercluster, Shapley Supercluster, and Horologium-Reticulum Supercluster.  Note that we mark multiple overdensities for a single structure in two cases:  We mark both of the two main overdensities of the Shapley Supercluster (as identified by \citealt{fairall98}), and four overdensities of the Sculptor Wall.  The figure is indicative only, since superclusters are extended objects
and we have marked them as points.  Nonetheless, the points marked on the figure
provide a useful guide in the discussion in Section 5.1.2, in which we will compare
the locations of these superclusters to the features of the observed and
predicted velocity fields.

A few points that should be made in examining this comparison of the two models: 1) The Shapley Supercluster appears as the most prominent overdensity within $\sim 150$ \mpc, particularly in the 2MRS model.  2) There is no one particular region of the sky that shows an unusually strong deviation between the two models.  Rather, the deviations between the models are scattered across the sky, with the largest differences appearing on the outskirts of the survey volume  3) The plot illustrates what was stated in the previous subsection: The scatter in densities (and velocities) is somewhat larger in the 2MRS model than the PSCz model, meaning that the former model includes more overdense superclusters and more underdense voids.  This may suggest that the fiducial value of $\beta$ that was assumed for the 2MRS model is too large, or that the fiducial value of $\beta$ assumed for the PSCz model is too small.  We explore that possibility in greater detail in a future paper (Magoulas et al. in prep.).

While Figure 3 shows that the familiar features of large-scale structure are apparent in both models, we can also ask whether the two model density fields predict similar peculiar velocities on the scale of individual gridpoints. In Figure 4, we show contour plots of 2MRS model velocities plotted against PSCz model velocities for all gridpoints out to 16,120 \kms the redshift limit of 6dFGSv. For nearby points ($cz < 8000$ \kms) the velocities in the two models are, as expected, positively correlated with a slope close to unity. However at larger distances the correlation grows weaker, mainly because the number of galaxies is decreasing rapidly with redshift for both surveys (again, see the comparison in \citealt{erdogdu06}).  At redshifts of $\sim12,000$ \kms or greater (where roughly half of our 6dFGSv galaxies lie), there is virtually no correlation between the velocities of the two models on the scale of individual grid points. This is because the PSCz survey in particular has very sparse sampling at these distances and so is heavily smoothed.

\begin{figure*} \centering 
\begin{minipage}{175mm}
\includegraphics[width=1.0\textwidth]{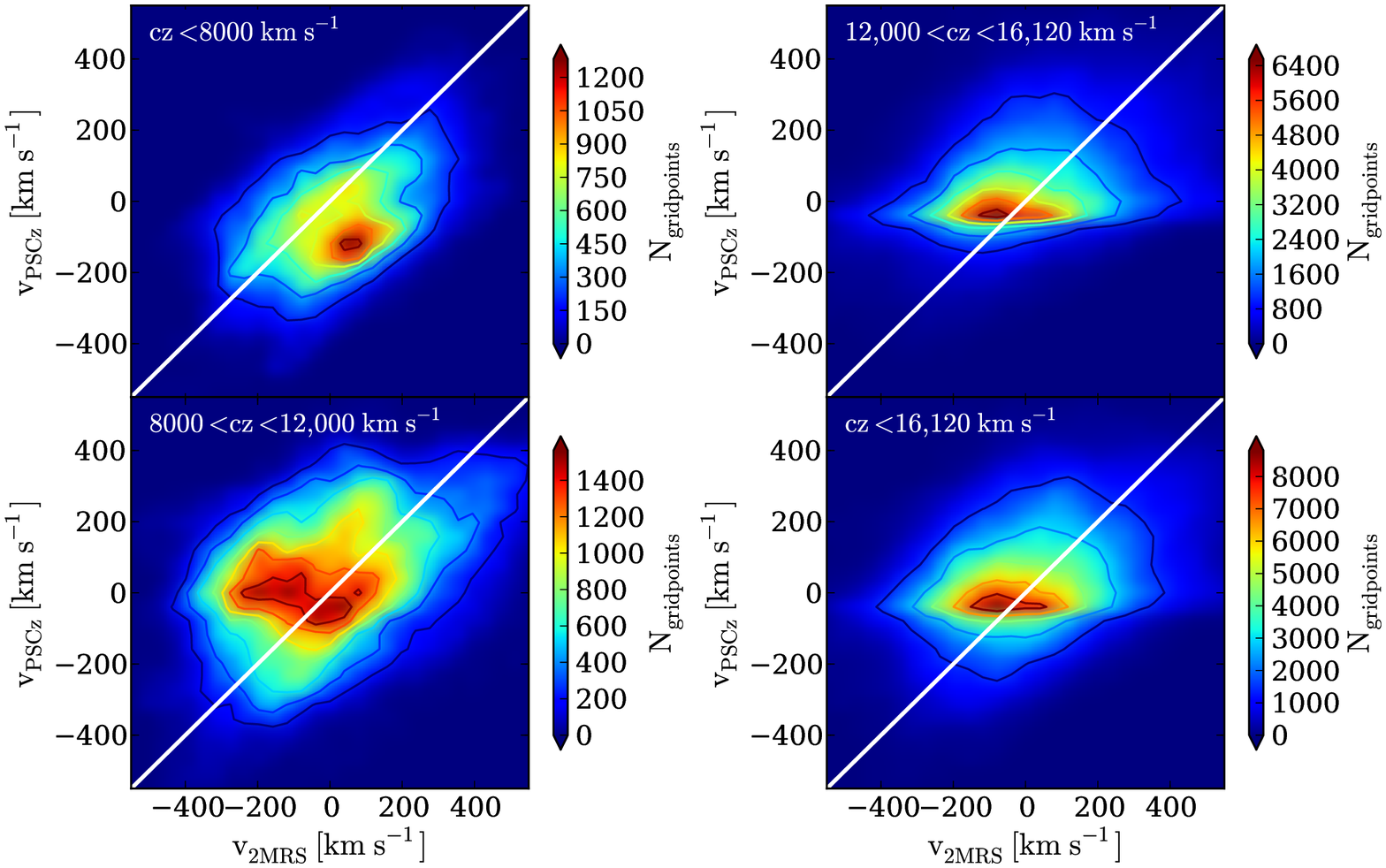}
\caption{Contour plots, comparing the model velocities from 2MRS and PSCz, calculated at individual gridpoints for different redshift slices.  The lower right panel shows all gridpoints with $cz<16,120$ \kms, which is the redshift limit of 6dFGSv.  The remaining three panels show subsets of this volume, with redshift ranges written at the top of each panel.  The colorbars show the number of gridpoints found within a single smoothing length, which corresponds to 40 \kms $\times$ 40 \kms.  The white diagonal line in each panel shows the 1-to-1 line.  We note that while the velocities from the two models appear to be correlated at low redshift, that correlation fades way for the gridpoints at higher redshift (where we find the bulk of our 6dFGSv galaxies).
\label{FIG4}} 
\end{minipage} 
\end{figure*}

\section{Fitting the Fundamental Plane}

\subsection{Maximum likelihood methodology}

We employ a maximum likelihood method to fit the FP,  similar to the method
developed by \citet{colless01a} and
\citet{saglia01} to fit the EFAR sample. The method is explained in detail in
\citet{magoulas12x}, but summarized below.

As \citet{colless01a} noted, when plotted in $r$-$s$-$i$ space,
early-type galaxies are well represented by a 3D Gaussian distribution.  This was shown to be true for 6dFGS by \citet{magoulas12x}.  (See, e.g., Figure 9 from that paper.)  Our maximum likelihood method then involves fitting the distribution of
galaxies in $r$-$s$-$i$ space to a 3D Gaussian, where the shortest axis
is orthogonal to the FP and characterises the scatter about the plane,
while the other two axes fit the distribution of galaxies within the
plane.

For this functional form, the probability density $P({\bf x_n})$ of observing the $n$th
galaxy at FP space position ${\bf x_n}$ can be computed according to \citet{magoulas12x}
equation~4,
\be 
P({\bf x_{n}}) = \frac{ \exp[-\frac{1}{2}{\bf x^{T}_{n}}({\bf \Sigma} +
  {\bf E_{n}})^{-1}{\bf x_{n}}]}{(2\pi)^{\frac{3}{2}}|{\bf \Sigma} +
  {\bf E_{n}}|^{\frac{1}{2}}f_{n}} ~, 
\ee
where ${\bf \Sigma}$ is the variance matrix for the 3D Gaussian
describing the galaxy distribution, ${\bf E_{n}}$ is the
observational error matrix, $\bf{ x_{n}}$ is the position in FP space
given by $(r-\bar{r}, s-\bar{s}, i-\bar{\imath})$, and $f_n$ is a
normalisation term depending on the sample selection function.  (Quantities with $n$ subscripts are specific to the particular galaxy.)  In
logarithmic form, this is 
\begin{align} 
  \ln(P({\bf x_{n}})) = &-[\frac{3}{2} \ln(2\pi) + \ln(f_{n}) +
                       \frac{1}{2}\ln(|\mathbf{\Sigma}+\mathbf{E_{n}}|)
                       \notag \\  
                      &+\frac{1}{2}\mathbf{x_{n}^{T}}(\mathbf{\Sigma} +
                       \mathbf{E_{n}})^{-1}\mathbf{x_{n}}] ~.
\end{align}

The intrinsic 3D Gaussian distribution of galaxies in FP space is defined by the variance matrix ${\bf \Sigma}$, which has eight parameters: $a$ and $b$ (which determine its orientation), $\bar{r}$, $\bar{s}$, and
$\bar{\imath}$ (which set the centroid), and $\sigma_1$, $\sigma_2$, and $\sigma_3$ (which determine its extent), as given by the relations provided by \citet{magoulas12x}.  Our maximum likelihood fitting method involves finding the
values of the eight fitted parameters that maximize the total
likelihood,
\be 
\ln(\mathcal{L}) = \sum_n \ln(P({\bf x_{n}})) ~. 
\ee
This is achieved by searching the parameter space with a non-derivative
multi-dimensional optimization algorithm called BOBYQA (Bound
Optimization BY Quadratic Approximation; \citealt{powell06a}).

The method is described in detail in \citet{magoulas12x}.  However, as explained in Section 2.1 of this paper, the catalog has been revised since that paper was published.  As a result, the fitting method has been applied to the revised catalog, which yields a best fit 6dFGS $J$-band FP of
\be
r = (1.438{\pm}0.023)s + (-0.887{\pm}0.008)i + (-0.108{\pm}0.047) 
\ee
where $r$, $s$ and $i$ are in units of log[$h^{-1}$\,kpc], log[\kms],
and log[$L_{\odot}$\,pc$^{-2}$] respectively. For converting between physical and
angular units, we assume a flat cosmology with $\Omega_m=0.3$ and
$\Omega_\Lambda =0.7$, though the specifics of the assumed cosmology
affect the FP fit very weakly.

We should also note that in previous papers, we investigated the possibility of adding one or more additional parameters to the FP fit.  \citet{magoulas12x} investigated FP trends with such parameters as environment and morphology.  \citet{springob12} investigated the FP space trends with stellar population parameters.  The only supplemental parameter with the potential to improve our FP fit was stellar age, as explained by \citet{springob12}.  However, as stated in that paper, while stellar age increases the scatter of the FP, the uncertainties on the measured ages of individual galaxies are too large to allow useful corrections for galaxy distances.  We thus do not include any corrections for stellar age, or any other stellar population parameters, in the FP fitting done here.

\subsection{Calibrating the FP zeropoint}

In the expression $r=as+bi+c$, the value of $c$ gives us the zeropoint, and the calibration of the relative sizes of the galaxies depends on how one determines the value of $c$.  As we are using the FP to measure peculiar velocities, it also gives us the zeropoint of peculiar velocities (or more specifically, the zeropoint of logarithmic distances).  We need to make some assumption about the peculiar velocity field, in order to set the zeropoint of the relation.  When we fit the FP
in \citet{magoulas12x}, for example, we assumed that the average radial
peculiar velocity of the galaxies is zero.

For a large all-sky sample, the assumption that the average peculiar
velocity of the sample is zero is equivalent to assuming that the
velocity field includes no monopole term, since a monopole (an error in
the expansion rate given by $H_0$) is completely degenerate with an
offset in the FP zeropoint. As long as the galaxies are evenly
distributed across the whole sky, this assumption only impacts the
velocity field monopole, and we can still measure higher order
multipoles. The situation is somewhat different in the case of a sample
that only includes galaxies in one hemisphere, however. In this case,
assuming that the mean velocity of the galaxies in the sample is zero
means suppressing the polar component of the dipole. In the case of a southern hemisphere survey with a dipole motion directed
along the polar axis, for example, assuming that the mean velocity of the galaxies is
zero corresponds to calibrating out the entire dipole.

6dFGS is of course just such a hemispheric survey. We can essentially
eliminate this FP calibration problem, however, if we set the zeropoint
using only galaxies close to the celestial equator. Such a `great
circle' sample is, like a full-sphere sample, only degenerate in the
monopole---even if the velocity field includes a dipole with a large
component along the polar direction, this component has negligible impact on
the radial velocities of galaxies close to the celestial equator.

How do we then set the zeropoint with such a great circle sample?  Recall that the zeropoint $c$ is not represented by any of the 8 parameters in our 3D Gaussian model alone.  Rather, it is a function of multiple FP parameters (explicitly, $c=\bar{r}-a \bar{s}-b \bar{i}$).  However, if we fix the other FP parameters to the fitted values for the entirety of the sample, then the zeropoint is represented by $\bar{r}$.  This is a quantity that we measured to be $\bar{r}=0.184 \pm 0.005$ for the whole sample.  However, because of degeneracies between $\bar{r}$, $\bar{s}$, $\bar{i}$, and the slopes $a$ and $b$, 0.005 dex does not represent the true uncertainty in the zeropoint.

We define our equatorial sample as the $N_g = 3781$ galaxies with
$-20^\circ\leq\delta\leq0^\circ$, and fit the FP zeropoint ($\bar{r}$)
after fixing the other coefficients that define the FP
($a,b,\bar{s},\bar{\imath},\sigma_1,\sigma_2, \sigma_3$) to the best-fit
values from the full sample.  The best-fit value of the mean effective
radius for the equatorial subsample is $\bar{r}=0.178$.  For this equatorial sample, the uncertainty on $\bar{r}$ would be 0.007 dex if we were fitting for all 8 FP parameters.  However, because we constrain all parameters except for $\bar{r}$, the uncertainty is only 0.003 dex, and this represents our uncertainty in the zeropoint of the relation.  This measurement of the uncertainty assumes a Gaussian distribution of peculiar velocities in the great circle region, with no spatial correlation.

Despite the fact that we have calibrated the zeropoint of the FP relation using the galaxies in the range $-20^\circ\leq\delta\leq0^\circ$, to mitigate the possibility of a large {\it dipole} motion biasing the zeropoint, we cannot rule out the possibility of a {\it monopole} in the velocity field within that volume creating such a bias.  In Section 5.1.2, we explore the possibility of an offset in the zeropoint of the FP relation in greater detail.

Finally, we note that there is the potential for some bias in the zeropoint due to the fact that $-20^\circ\leq\delta\leq0^\circ$ is not exactly a great circle.  The extent of this bias depends on the size of the bulk flow's polar component relative to the mean velocity dispersion of galaxies.  If we assume that the mean velocity dispersion of galaxies is $\sim 300$ \kms, then we estimate that a comparably large bulk flow of $\sim 300$ \kms along the celestial pole would introduce a bias of 0.0007 dex in the zeropoint.  This is much smaller than our 0.003 dex uncertainty, however, and a $\sim 300$ \kms bulk flow along the polar direction alone is most likely a pessimistically large estimate.

\section{Deriving peculiar velocities}

\subsection{Bayesian distance estimation}

When measuring a galaxy distance, authors typically derive a single
number, along with its error. If the distance
estimates have a Gaussian distribution, these two numbers fully
characterise the probability distribution for the galaxy distance.
However, it is often more natural to estimate the logarithm of the
distance, particularly if it is this quantity that has a Gaussian error distribution. 
This in fact applies to distance estimates using the
FP, which is fit in logarithmic $r$-$s$-$i$ space.

However, because of the various selection effects and bias corrections, the
probability distribution is not exactly Gaussian in logarithmic
distance. Thus in order to retain all the available information, we
choose to calculate the full posterior probability distributions for the
distance to each galaxy. This requires that we have a clear
understanding of the definitions being used in the previous section.

We have referred to `Fundamental Plane space', by which we mean the 3D
parameter space defined by $r$, $s$ and $i$.  $r$, $s$
and $i$ can be described either as observational parameters or physical parameters. That
is, galaxy radius, velocity dispersion, and surface brightness are all
clearly observational parameters, but they are also (when defined
appropriately) physical properties that the galaxy possesses independent
of any particular set of observations. When we fit the FP we are
simultaneously fitting an empirical scaling relation of observable
quantities for our particular sample and deriving a scaling relation of
physical quantities that should hold for any similarly-selected sample.

There is, however, a distinction that needs to be made for $r$. The
observed quantity is actually $r_z$, or the physical radius (in
logarithmic units) that the galaxy would have if it was at its redshift
distance. (The actual observables here are angular radius and redshift,
but $r_z$ is a convenient and well-defined function of those
observables.) Using the definition of angular diameter distance $d^A$
(in logarithmic units), we have $r_z-r_H=d^A_z-d^A_H$, where $d^A_z$ and
$d^A_H$ are the logarithms of the angular diameter distances corresponding (respectively) to
the observed redshift of the galaxy and the Hubble redshift (cf.\
equation~2 and \citealt{colless01a} equation~8). However, the relevant
distance for our purposes (i.e.,\ for measuring peculiar velocities) is
logarithmic comoving distance, $d$, which is related to logarithmic
angular diameter distance by $d=d^A+\log(1+z)$. Hence 
\be 
r_z-r_H = d_z-d_H-\log [(1+z)/(1+z_H)]
\ee 
where $z$ is the observed redshift and
$z_H$ is the redshift corresponding to the Hubble distance of the
galaxy.  The $\log[(1+z)/(1+z_H)]$ term thus accounts for the difference between angular
diameter distance and comoving distance.  At this point, we define the shorthand
\be
\Delta r=r_z-r_H
\ee
\be
\Delta d=d_z-d_H
\ee
\be
\Delta z=\log [(1+z)/(1+z_H)]
\ee
Our goal is to derive $P(d_{H,n}|r_{z,n},s_n,i_n)$, which is the
probability distribution of the $n$th galaxy's comoving distance $d_H$,
given the observational parameters $r_z$ (the galaxy's size, assuming it
is at the distance corresponding to its observed redshift),
$s$, and $i$.  For any given
galaxy, $r_z$ and $d_z$ are observed directly, but $r_H$ and $d_H$ must be
determined.

How then do we calculate the probability distribution for distance?
Because equation~5 provides the probability distribution of physical
radius for given values of velocity dispersion and surface brightness,
the simplest approach available to us is to calculate
$P(\Delta r_{n}|r_{z,n},s_n,i_n)$ over an appropriate range of $\Delta r$ values, and then use a transformation of variables to get
$P(\Delta d_{n}|r_{z,n},s_n,i_n)$.
$P(\Delta r_{n}|r_{z,n},s_n,i_n)$ is the posterior probability that
the ratio of the $n$th galaxy's size at its redshift distance
to its size at its true comoving distance (in logarithmic
units) is $\Delta r_{n}$. $P(\Delta d_{n}|r_{z,n},s_n,i_n)$ is
the corresponding posterior probability for the ratio of comoving
distances for galaxy $n$. Since the redshift of the galaxy is
given and $d_z$ is known, $P(\Delta d_{n}|r_{z,n},s_n,i_n)$ is
equivalent to $P(d_{H,n}|r_{z,n},s_n,i_n)$.

We implement this approach in the following manner:

(1) Specify the FP template relation using our fitted 3D Gaussian model,
as described by the eight parameters $a$, $b$, $\bar{r}$, $\bar{s}$,
$\bar{\imath}$, $\sigma_1$, $\sigma_2$, and $\sigma_3$. The best-fit
values of these parameters are given in \citet{magoulas12x} for various
samples and passbands. For the full $J$-band sample that we are using
here, the best-fit values are: $a=1.438$, $b=-0.887$, $\bar{r}=0.178$,
$\bar{s}=2.187$, $\bar{\imath}=3.175$, $\sigma_1=0.047$,
$\sigma_2=0.315$, and $\sigma_3=0.177$.  The value of $\bar{r}$ was specifically fit to the region $-20^\circ\leq\delta\leq0^\circ$, as explained in Section 3.2.

(2) For each individual galaxy $n$, loop through every possible
logarithmic comoving distance $d_{H,n}$ that the galaxy could have.
Distance is of course a continuous quantity, but in practice we are
limited to examining a finite number of possible distances. We consider 501 evenly spaced values of
$\Delta d_{n}$, between -1.0 and +1.0 in steps of 0.004 dex, and
compute the corresponding values of $\Delta r_{n}$.  These steps correspond to $1\%$ in relative distance.

(3) For each of these possible logarithmic ratios of radius, use Bayes'
theorem to obtain the posterior distribution for the $n$th galaxy's size
as a function of the observables,
\be
P(\Delta r_{n}|r_{z,n},s_n,i_n) = {P(r_{z,n},s_n,i_n|\Delta r_{n}) P(\Delta r_{n}) 
   \over P(r_{z,n},s_n,i_n)} 
\ee
Given our assumed physical radius $r_{H,n}$, we can evaluate
$P({\bf x_{n}})$ in equation~5, on the assumption that $P({\bf x_{n}}) =
P(r_{z,n},s_n,i_n|\Delta r_{n})$, so long as the ${\bf x_{n}}$ in
question uses the physical radius corresponding to the distance
$d_{H,n}$.  That is, while equation~5 is written in such a way
that it suggests that there is a single probability density $P({\bf x_n})$ for
galaxy $n$, we now suggest that for galaxy $n$, we must consider many
possible distances that the galaxy could be at, each of which
corresponds to a different radius and different ${\bf x_n}$.

Having evaluated $P(r_{z,n},s_n,i_n|\Delta r_{n})$, we multiply by
the prior, $P(\Delta r_{n})$, assumed to be flat, and apply the proper normalization (that is, normalizing the
probabilities so that the total probability across all possible radii is
unity), to give us the posterior probability
$P(\Delta r_{n}|r_{z,n},s_n,i_n)$. 
 
(4) Convert the posterior probability distribution of sizes,
$P(\Delta r_{n}|r_{z,n},s_n,i_n)$, to that of distances,
$P(\Delta d_{n}|r_{z,n},s_n,i_n)$, by changing variables from $r$
to $d$. To do so, we use the fact that
\be
P(\Delta d_{n}) = P(\Delta r_{n}) {d[\Delta r_{n}] 
                       \over d[\Delta d_{n}]}
\ee

Let us now define $D_{z,n}$ and $D_{H,n}$ as the linear comoving distances of the galaxy $n$ in units of \mpc, corresponding (respectively) to the observed redshift and the Hubble redshift of the galaxy.  That is, they are the linear equivalents of the logarithmic $d_{z,n}$ and $d_{H,n}$.  From the chain rule, we have
\be
{d\Delta r \over d\Delta d} = 1-{d\Delta z \over d\Delta d} = 1-{d\Delta z \over dz_H}{dz_H \over dD_H}{dD_H \over d\Delta d}
\ee

$d\Delta z / dz_H$ and $dD_H / d\Delta d$ can be evaluated relatively easily.  However, in order to evaluate $dD_H / d\Delta d$, we must examine the relationship between redshift
and comoving distance.  Assuming a standard $\Lambda$CDM cosmology with $\Omega_m=0.3$ and
$\Omega_\Lambda =0.7$, we numerically integrate the relations given by \citet{hogg99}, to get the following low redshift approximation, relating the redshift in \kms to the comoving distance in \mpc:
\be 
cz \approx k_1 D_H + k_2 D_H^2
\ee
where $k_1=99.939$ and $k_2=0.00818$.  Evaluating the relevant derivatives gives us
\be 
P(\Delta d_{n}) \approx P(\Delta r_{n}) 
        (1-{k_1 D_H + 2k_2 D_H^2 \over c (1+z_H)})
\ee
with $c$ expressed in units of \kms.  We use this numerical approximation in computing the peculiar velocities for the 6dFGS galaxies, as it is extremely accurate over the range of redshifts of interest. However we note that the approximate formula of \citet{lynden-bell88} also provides adequate precision and can be used with different cosmological models through its dependence on $q_0$.

The question of how to calculate the different normalization terms
$f_{n}$ in equation~5 is addressed in the following subsection.
However, it should be noted that whether one needs to include this term
at all depends on what precisely the probability distribution in
question is meant to represent. If we were interested in the probability
distribution of possible distances for each individual galaxy considered
in isolation, then the $f_{n}$ term should be omitted.  In this case, however, we are computing the
probability distribution of the comoving distance corresponding to the
redshift-space position of galaxy $n$, and so the $f_{n}$ term must be included.

\subsection{Selection bias}

Malmquist bias is the term used to describe biases originating from the
spatial distribution of objects \citep{malmquist24}. It results from the
coupling between the random distance errors and the apparent density distribution
along the line of sight. There are two types of distance errors that one
must consider. The first of these is inhomogeneous Malmquist bias, which
arises from local density variations due to large-scale structure along
the line of sight. It is most pronounced when one is measuring galaxy
distances in real space.  This is because the large distance errors cause one to observe galaxies scattering away from overdense regions, creating artificially inflated measurements of infall onto large structures.  By contrast, when the measurement is done in redshift space,
the much smaller redshift errors mean that this effect tends to be negligible (see e.g., \citealt{strauss95}).

For the 6dFGSv velocity field as presented in this paper, we
are measuring galaxy distances and peculiar velocities in redshift space
rather than real space. In this case, inhomogeneous Malmquist bias is negligible, and
the form of Malmquist bias that we
must deal with is of the second type, known as homogeneous Malmquist
bias, which affects all galaxies independently of their position on the
sky. It is a consequence of both (1)~the volume effect, which means that
more volume is covered within a given solid angle at larger distances
than at smaller distances, and (2) the selection effects, which cause
galaxies of different luminosities, radii, velocity dispersions etc., to
be observed with varying levels of completeness at different distances.  We note, however, that different authors use somewhat different terminology, and the latter effect described above is often simply described as `selection bias'.

The approach one takes in correcting for this bias
depends in part on the selection effects of the survey. If the selection
effects are not well defined analytically, then the bias
correction can be complicated, though still possible.  For example, \citet{freudling95x} use Monte Carlo simulations to
correct for Malmquist bias in the SFI sample (\citealt{giovanelli94}, \citealt{giovanelli95}).

In our case, however, we have explicit analytical expressions for the
intrinsic distribution of physical parameters, and explicit and
well-defined selection criteria. It is thus possible, at least in
principle, to correct for selection bias analytically.  However, as we
will see, in practice we are obliged to use mock samples for the purposes
of evaluating the relevant integral.

Our bias correction involves applying an appropriate weighting
to each possible distance that the galaxy could be at in order to
account for galaxies at those distances that are not included in our
sample due to our selection criteria. One complication is that certain
regions of FP space are not observed in our sample because of our source selection. The expression for the likelihood that we give in Equation~5
includes a normalization factor $f_{n}$ that ensures the integral of
$P({\bf x_{n}})$ over all of FP space remains unity, even when certain
regions of FP space are excluded by selection cuts.

Let us consider one such selection effect. Suppose there is a lower
limit on $s$, which we call $s_{cut}$, above which we observe all
galaxies and below which we observe none. Then $P({\bf x_{n}})=0$ for
$s<s_{cut}$, and $P({\bf x_{n}})$ follows Equation~5 for $s \geq s_{cut}$.
We must include the normalization factor $f_{n}$ here, which in this
case is
\be 
f_{n} = \int_{-\infty}^\infty \int_{s_{cut}}^\infty \int_{-\infty}^\infty 
          {exp[-0.5 {\bf x_{n}^T} ({\bf \Sigma} + {\bf E_n} )^{-1} {\bf x_{n}}]
          \over (2\pi )^{1.5} |{\bf \Sigma} + {\bf E_n} |^{0.5}}\,di\,ds\,dr
\ee
In practice, because $P(d_{H,n}|r_{z,n},s_n,i_n)$ is normalized to 1 (across the range of $d_{H,n}$ values for each galaxy), the $f_{n}$ only comes into play if it varies for different distances.  It thus turns out to be irrelevant in this case, because $s$ is distance-independent and $s_{cut}$ does not change as a
function of galaxy distance or peculiar velocity.

We next consider what happens when, in addition to the $s$ cut, we also have an apparent magnitude cut. At a particular
logarithmic distance $d$, this corresponds to a cut in absolute magnitude,
$M_{cut}(d)$. At that distance, we observe all galaxies with $s$ greater
than $s_{cut}$ and $M$ brighter than $M_{cut}(d)$, whereas we miss all
others. A cut in absolute magnitude corresponds
to a diagonal cut in $r$-$s$-$i$ space, since absolute magnitude is a
function of both $r$ and $i$.  We can incorporate this cut into the equation for $f_{n}$ by integrating $i$ from $-\infty$ to $\infty$, but $r$ from $r_{cut}(i)$ to $\infty$, where $r_{cut}(i)$ is the radius at the surface brightness $i$ corresponding to $M_{cut}$.

We can then rewrite the expression for $f_{n}$ as
\be 
f_{n} = \int_{r_{cut}(i)}^\infty \int_{s_{cut}}^\infty \int_{-\infty}^\infty 
          {exp[-0.5 {\bf x_{n}^T} ({\bf \Sigma} + {\bf E_n} )^{-1} {\bf x_{n}}]
          \over (2\pi )^{1.5} |{\bf \Sigma} + {\bf E_n} |^{0.5}}\,di\,ds\,dr
\ee

Unfortunately, there is no easy way to evaluate this integral
analytically. We thus determine $f_{n}$ using a large Monte Carlo
simulation of a FP galaxy sample (with $N_g = 10^5$) drawn from the
best-fit $J$-band FP values and our 6dFGS selection function. The entire
mock sample of galaxies is used to calculate the value of $f_{n}$ as a function of distance, as seen in Figure 5.

\begin{figure}
\includegraphics[width=0.45\textwidth]{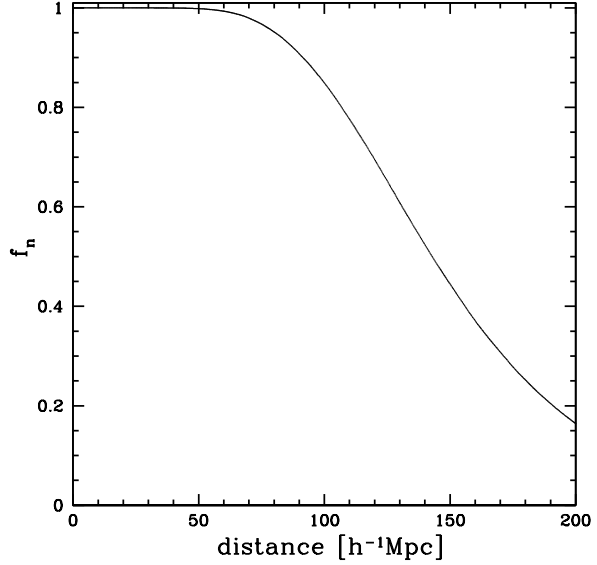}
\caption{The normalization factor used to correct selection
  bias as a function of distance, derived from a mock sample with
  100,000 galaxies.
\label{FIG5}}
\end{figure}

Note that each galaxy has its own individual error matrix, ${\bf E_n}$, and we should be using the specific ${\bf E_n}$ matrix for galaxy $n$.  However, running such an $N_g = 10^5$ Monte Carlo simulation separately for all $\sim 9000$ galaxies is computationally impractical.  As a compromise, when we run the Monte Carlo simulation, we assign measurement errors to every mock galaxy parameter according to the same algorithm specified for 6dFGS mock catalogs explained in \citet{magoulas12x} Section 4.  This treats the mock galaxy measurement errors as a function of apparent magnitude.

\subsection{Peculiar Velocity Probability Distributions}

In Figure~6 we show the posterior probability density distributions as
functions of logarithmic distance for ten randomly-chosen 6dFGSv
galaxies.

\begin{figure}
\includegraphics[width=0.45\textwidth]{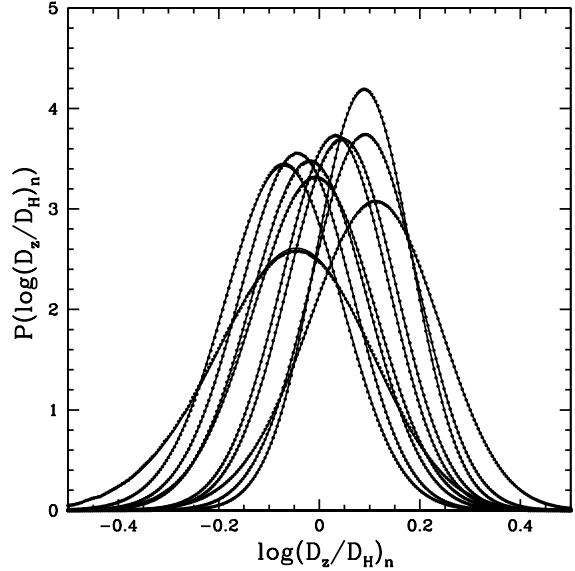}
\caption{For ten randomly-chosen galaxies in 6dFGSv, we show the
  probability density distribution of $\Delta d_{n}=\log(D_z / D_H)_n$, which is the
  logarithm of the ratio of the comoving distance associated with galaxy
  $n$'s redshift to the true comoving distance of the galaxy.  The exact probability distributions are represented by circles, whereas the approximations from Equation 21 are represented by solid lines.
\label{FIG6}}
\end{figure}

Because the probability distributions are nearly Gaussian, we fit Gaussian functions to the distribution for each galaxy, and calculate the mean value $\langle \Delta d \rangle$ and the width of the Gaussian $\epsilon_d$ (and thus, the error on the logarithmic distance ratio).  While the skewness of the distributions is sufficiently small that ignoring it and assuming a simple Gaussian distribution should be adequate for most cosmological applications, we do also calculate the parameter $\alpha$ for each galaxy to characterize the skewness.  $\alpha$ describes the skewness according to the Gram-Charlier series (see, e.g., \citealt{cramer46}).  We begin with the standard Gaussian distribution
\be 
P(\Delta d)=G(\Delta d, \epsilon_d) = {e^{-(\Delta d - \langle \Delta d \rangle)^2 / 2 \epsilon_d^2} \over {\epsilon_d \sqrt{2\pi}}}
\ee
which is then modified to take the form
\be
P(\Delta d)=G(\Delta d, \epsilon_d) \left[{1+\alpha \left( \left({\Delta d - \langle \Delta d \rangle \over \epsilon_d}\right)^3 - {3 (\Delta d - \langle \Delta d \rangle ) \over \epsilon_d}\right)}\right]
\ee
To compute $\alpha$ for galaxy $n$, we evaluate $\alpha_{n,i}$ in the $i$th bin of $\Delta d$ for that galaxy, sampled across a subset of the same 501 evenly spaced values between -1.0 and +1.0 that are described in Section 4.1:
\begin{align}
\alpha_{n,i}=&\left[{{PDF(\Delta d_{n,i}) \over G(\Delta d_{n,i}, \epsilon_{d,n})} -1}\right]
                    \notag \\
                    &\left[{\Delta d_{n,i} - \langle \Delta d_n \rangle \over \epsilon_{d,n}}^3 - {3 (\Delta d_{n,i} - \langle \Delta d_n \rangle ) \over \epsilon_{d,n}}\right]
\end{align}
where $PDF(\Delta d_{n,i})$ is the probability density at $\Delta d_{n,i}$ for galaxy $n$ as described in Section 4.1, with the selection bias correction applied as in Section 4.2.  This is calculated across the range $\langle \Delta d \rangle -2\epsilon_d < \Delta d < \langle \Delta d \rangle +2\epsilon_d$, but excluding the range $\langle \Delta d \rangle -0.1\epsilon_d < \Delta d < \langle \Delta d \rangle +0.1\epsilon_d$ because the function is undefined for $\Delta d = \langle \Delta d \rangle$.  The mean value of $\alpha$ is -0.012, and it has a $1\sigma$ scatter of 0.011.

The values of $\langle \Delta d \rangle$, $\epsilon_d$, and $\alpha$ are given in Table 1.  The interested reader can reconstruct the probability distributions from Equation 21.  However, note that this is an approximation, which breaks down in the wings of the distribution, as it can yield (physically impossible) negative values when the function approaches zero.  The reconstructed probability distributions from Equation 21 for the ten galaxies shown in Figure 6 are represented in that figure by solid lines.

\begin{table*}
  \caption{6dFGSv logarithmic distance ratios, and associated parameters.  The columns are as follows: (1) source name in 6dFGS catalogue; (2, 3) right ascension and declination (J2000); (4) individual galaxy redshift in the CMB reference frame; (5) group redshift in the CMB reference frame, in cases where the galaxy is in a group (set to -1 for galaxies not in groups); (6) group identification number (set to -1 for galaxies not in groups); (7) the logarithmic distance ratio $\langle \Delta d \rangle = \langle \log(D_z/D_H)\rangle$; (8) the error on the logarithmic distance ratio, $\epsilon_d$, derived by fitting a Gaussian function to the $\Delta d$ probability distribution; (9) the skew in the fit of the Gaussian function, $\alpha$, calculated using Equation 22.  The full version of this table is available as an ancillary file, and will also be made available on {\it www.6dfgs.net}.}
\label{tab:vel_data}
\begin{tabular}{crrrrrrrr}
\hline \hline
\multicolumn{1}{c}{6dFGS name} & 
\multicolumn{1}{c}{R.A.} &
\multicolumn{1}{c}{Dec.} &
\multicolumn{1}{c}{$cz_{gal}$} &
\multicolumn{1}{c}{$cz_{group}$} &
\multicolumn{1}{c}{group number} &
\multicolumn{1}{c}{$\langle \Delta d \rangle$} &
\multicolumn{1}{c}{$\epsilon_d$} &
\multicolumn{1}{c}{$\alpha$} \\

\multicolumn{1}{c}{} &
\multicolumn{1}{c}{[deg.]} &
\multicolumn{1}{c}{[deg.]} & 
\multicolumn{1}{c}{[\kms]} &
\multicolumn{1}{c}{[\kms]} &
\multicolumn{1}{c}{} &
\multicolumn{1}{c}{[dex]} &
\multicolumn{1}{c}{[dex]} &
\multicolumn{1}{c}{}\\

\multicolumn{1}{c}{(1)} &  
\multicolumn{1}{c}{(2)} & 
\multicolumn{1}{c}{(3)} &
\multicolumn{1}{c}{(4)} &
\multicolumn{1}{c}{(5)} & 
\multicolumn{1}{c}{(6)} &
\multicolumn{1}{c}{(7)} & 
\multicolumn{1}{c}{(8)} &
\multicolumn{1}{c}{(9)}\\
\hline
g0000144-765225  & 	 0.05985	  &  -76.8736  & 	 15941  & 	 -1  & 	 -1  & 	 +0.1039  & 	 0.1296  & 	 -0.0200\\
g0000222-013746	   & 0.09225	  &  -1.6295	   & 11123	   & -1  & 	 -1	   & +0.0870	   & 0.0954	   & -0.0066\\
g0000235-065610  & 	 0.09780  & 	 -6.9362	  &  10920	   & -1	  &  -1	  &  +0.0282	  &  0.1073	  &  -0.0116\\
g0000251-260240  & 	 0.10455  & 	 -26.0445	  &  14926	   & -1	   & -1	  &  +0.0871	  &  0.1065	  &  -0.0111\\
g0000356-014547  & 	 0.14850  & 	 -1.7632	  &  6956	   & -1	   & -1	   & -0.0743	   & 0.1165	   & -0.0112\\
g0000358-403432  & 	 0.14895  & 	 -40.5756	  &  14746	   & -1	   & -1	   & -0.0560	   & 0.1532	   & -0.0217\\
g0000428-721715  & 	 0.17835	  &  -72.2874	  &  10366	   & -1	   & -1	   & -0.0486	   & 0.1123	   & -0.0135\\
g0000459-815803	  &  0.19125	  &  -81.9674	  &  12646	   & -1	   & -1	   & -0.0131	   & 0.1201	   & -0.0153\\
g0000523-355037	  &  0.21810	  &  -35.8437	  &  15324	   & 14646	   & 1261	   & -0.0219	   & 0.1145	   & -0.0083\\
g0000532-355911	  &  0.22155	   & -35.9863	  &  14725	   & 14646	   & 1261	   & +0.0421	   & 0.1077	   & -0.0098\\
\hline
\end{tabular}
\end{table*}

Note that while we use the group redshift for galaxies found in groups, we provide here the individual galaxy redshifts in Table 1 as well.  As explained in Section 2.1, we refer the interested reader to \citet{magoulas12x} for a more detailed description of the grouping algorithm.

In Figure 7, we show the histogram of the probability-weighted mean
values of the logarithm of the ratio of redshift distance to Hubble distance $\Delta d_{n}$ for each of the 8885 galaxies in the
6dFGSv sample; put another way, this is the histogram of expectation
values $\langle \Delta d_{n} \rangle$. The mean of this
distribution is $+$0.005\,dex, meaning that we find that the peculiar velocities in the survey volume are very slightly biased towards positive values.  The rms scatter is 0.112\,dex, which corresponds to an
rms distance error of 26\%. As explained in \citet{magoulas12x}, one
might naively assume that the 29\% scatter about the FP along the
$r$-axis translates into a 29\% distance error, but this neglects the
fact that the 3D Gaussian distribution of galaxies in FP space is not
maximized on the FP itself at fixed $s$ and $i$. The distance error
calculated by \citet{magoulas12x} neglecting selection bias is 23\%, but
the bias correction increases the scatter to 26\%.

\begin{figure}
\includegraphics[width=0.45\textwidth]{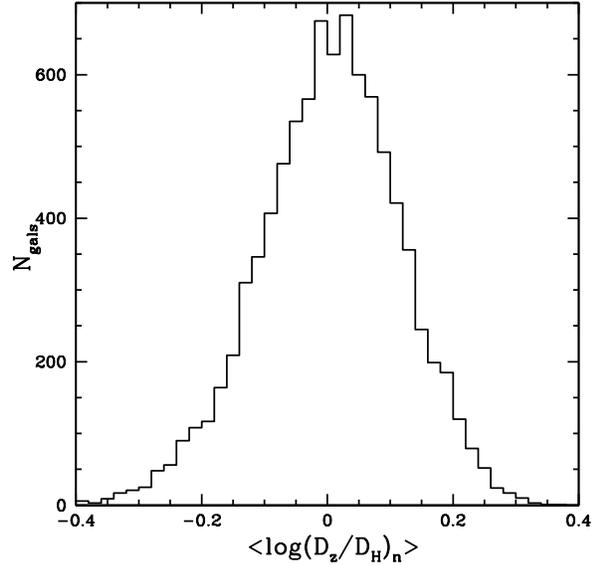}
\caption{Distribution of $\langle \Delta d_{n} \rangle = \langle \log(D_z / D_H)_n \rangle$, the expectation values of the logarithm of the ratio of redshift distance to Hubble distance $\Delta d_{n}$ for each of the 8885 galaxies in the 6dFGSv sample.
\label{FIG7}}
\end{figure}

We note that while all of our analysis is conducted in logarithmic distance units, some applications of the data may require conversion to linear peculiar velocities.  The interested reader is invited to convert these logarithmic distance ratios accordingly, accounting for the fact that the measurement errors are lognormal in peculiar velocity units.  Further elaboration on this point is provided in Appendix A.

\section{Peculiar Velocity Field Cosmography}

In a future paper we will perform a power spectrum analysis on the
peculiar velocity field in order to extract the full statistical
information encoded in the linear velocity field. In this
paper, however, we display the data in
such a way as to illuminate these correlations, and to give us a
cosmographic view of the velocity field. We approach this goal using
adaptive kernel smoothing.

We impose a 3D redshift-space grid in supergalactic cartesian
coordinates, with gridpoints 4\,$h^{-1}$\,Mpc apart. At each gridpoint
we compute adaptively smoothed velocities from both the 2MRS predicted
field and the 6dFGSv observed field using the following procedure.  It draws on methods used by \citet{silverman86} and \citet{ebeling06}, but we have adjusted these approaches slightly to produce smoothing kernels that, on average, lie in the range $\sim 5-10$ \mpc, as that appears to highlight the features of the velocity field around known features of large scale structure most effectively.

If $v({\mathbf r_i})$ is the logarithmic line-of-sight peculiar velocity
of gridpoint $i$ at redshift-space position ${\mathbf r_i}$, then our
smoothing algorithm defines $v({\mathbf r_i})$ according to the relation
\be 
v({\mathbf r_i}) = {\sum_{j=1}^{N_j} 
  v_j\cos\theta_{i,j}\,e^{-rr_{i,j}/2}\,\sigma_j^{-3} 
  \over \sum_{j=1}^{N_j} e^{-rr_{i,j}/2}\,\sigma_j^{-3} }
\ee
where: $\sigma_j$ is the
smoothing length of the 3D Gaussian kernel for galaxy $j$;
$\theta_{i,j}$ is the angle between the ${\mathbf r}$ vectors for the
gridpoint $i$ and galaxy $j$; and $rr_{i,j}$ is the square of the
distance between the gridpoint $i$ and galaxy $j$ in units of
$\sigma_j$. The index $j$ is over the $N_j$ galaxies in the sample for
which $rr_{i,j}<9$ (i.e.\ those galaxies within 3 smoothing lengths of
gridpoint $j$). 

The smoothing length $\sigma_j$ is defined to be a function of a
fiducial kernel $\sigma^\prime$ and a weighting depending on the local
density $\delta_j$
\be 
\sigma_j = 2\sigma^\prime\left[\frac{exp( \sum_{l=1}^{N}\ln\delta_l / N)}
           {\delta_j}\right]^\frac{1}{2}
\ee
where
\be 
\delta_j=\sum_{k=1}^{N_k} e^{-rr_{j,k}/2} 
\ee
and $rr_{j,k}$ is the square of the distance between galaxies $j$ and
$k$ in units of $\sigma^\prime$.  The summation on $k$ is over the $N_k$ galaxies within
3$\sigma^\prime$ of galaxy $j$, while the summation on $l$ is over all $N$ galaxies in the survey.  Thus the bracketed term in Equation 24 is the mean density for all galaxies divided by the local density $\delta_j$.  In our case, we set $\sigma^\prime =
10$\,$h^{-1}$\,Mpc, though we find that the actual smoothing length
$\sigma_j$ depends fairly weakly on the fiducial length $\sigma^\prime$.  The histogram of smoothing lengths is shown in Figure 8.
The mean smoothing length is $\langle\sigma_j\rangle=8.2$ \mpc.

\begin{figure}
\includegraphics[width=0.45\textwidth]{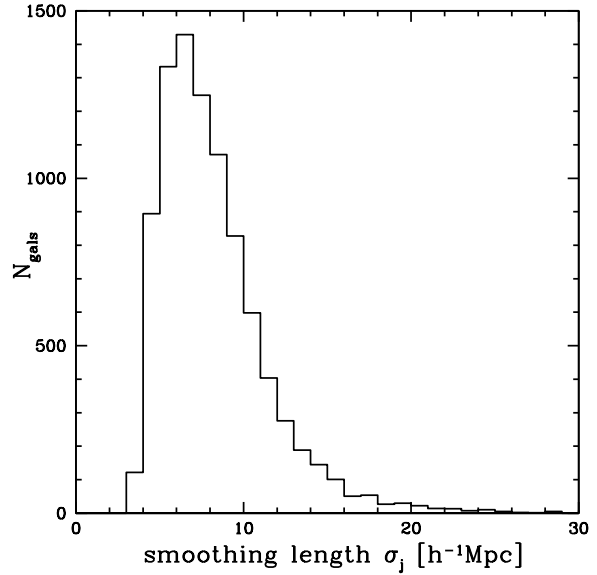}
\caption{Distribution of smoothing lengths, $\sigma_j$, for all 6dFGSv galaxies, following Equation 24.\label{FIG8}}
\end{figure}

\subsection{Features of the velocity field}

In Figures~9 and 10, we show the reconstructed 2MRS and PSCz velocity fields alongside the 6dFGS observed field.  In each case, the velocity field has been smoothed, using the adaptive kernel smoothing described above.  In Figure 9, the four panels on the left column show the smoothed velocity field predicted by 2MRS, in slices of SGZ.  The four panels in the central column show the observed 6dFGS velocity field, smoothed in the same manner.  The four panels in the right column show the difference between the 2MRS velocity field and the 6dFGS velocity field.  Figure 10 follows the same format, but with the PSCz field in place of 2MRS.  That is, the left column corresponds to the velocity field predicted by PSCz, and the right column corresponds to the difference between the PSCz field and the 6dFGS field.  In each case the color-coding gives the mean smoothed logarithmic distance ratio averaged
over SGZ at each (SGX,SGY) position.  We note that while Figure 4 showed that the correlation between the 2MRS and PSCz model velocities weakens at higher redshifts, we see in Figures 9 and 10 that both models make qualitatively similar predictions for the velocity field on large scales.

\begin{figure*} \centering 
\begin{minipage}{175mm}
\includegraphics[width=1.0\textwidth]{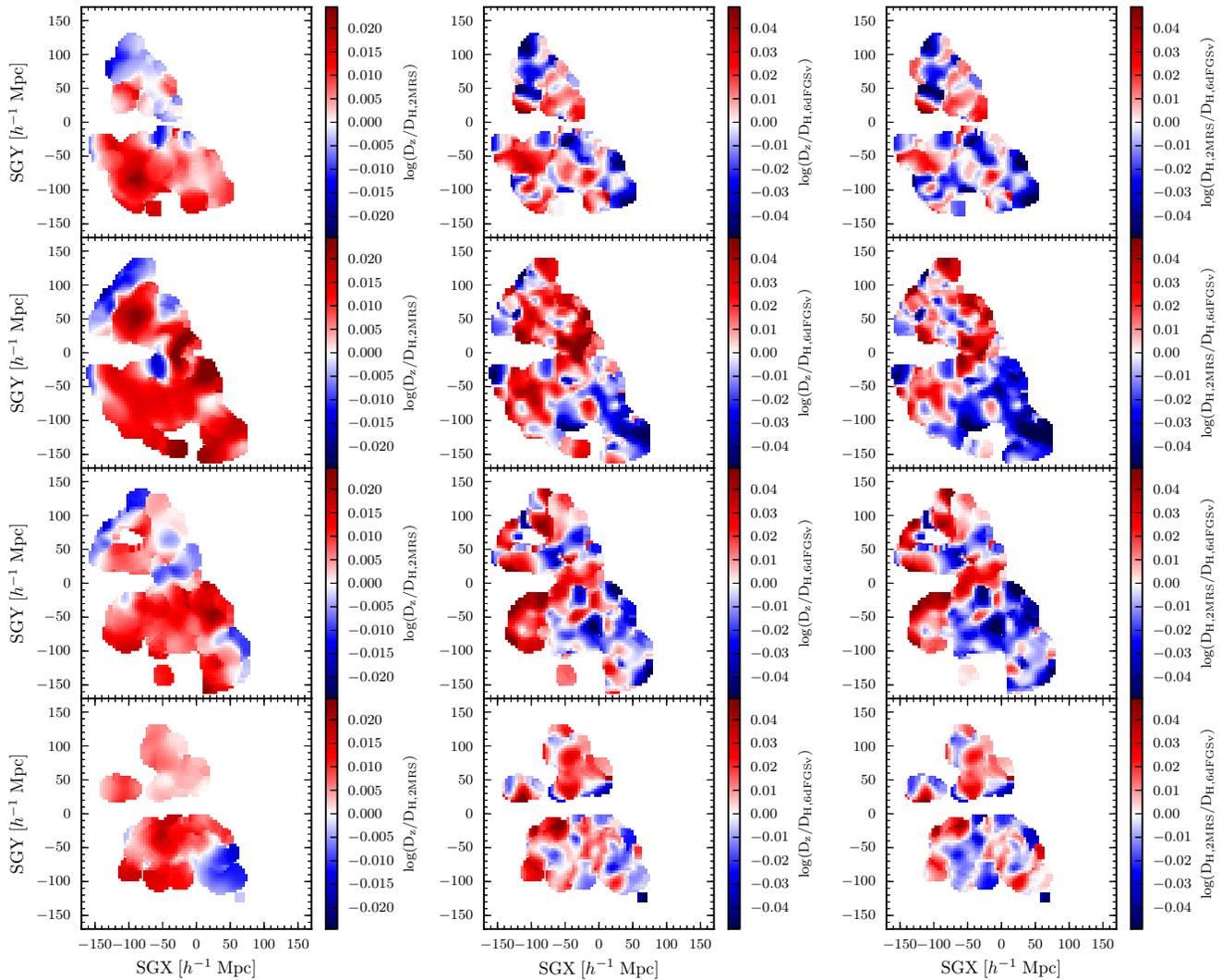} 
\caption{ Adaptively smoothed versions of the reconstructed 2MRS velocity field, as derived by Erdo{\u g}du et al. (submitted)
  (left), the observed 6dFGS velocity field (centre) and the observed
  6dFGS field minus the 2MRS reconstruction (right), in the same four slices of SGZ that are displayed in Figure 3.  In each case, the velocity field is given in logarithmic distance units ($\Delta d=\log(D_z / D_H)$, in the nomenclature of Section 4.1), as the logarithm of the ratio between the redshift distance and the true Hubble distance.  As shown in the colorbars for each panel, redder (bluer) colors correspond to more positive (negative) values of $\Delta d$, and thus more positive (negative) peculiar velocities.  Gridpoints are spaced 4 \mpc \ apart.\label{FIG9}} 
\end{minipage} 
\end{figure*}

\begin{figure*} \centering 
\begin{minipage}{175mm}
\includegraphics[width=1.0\textwidth]{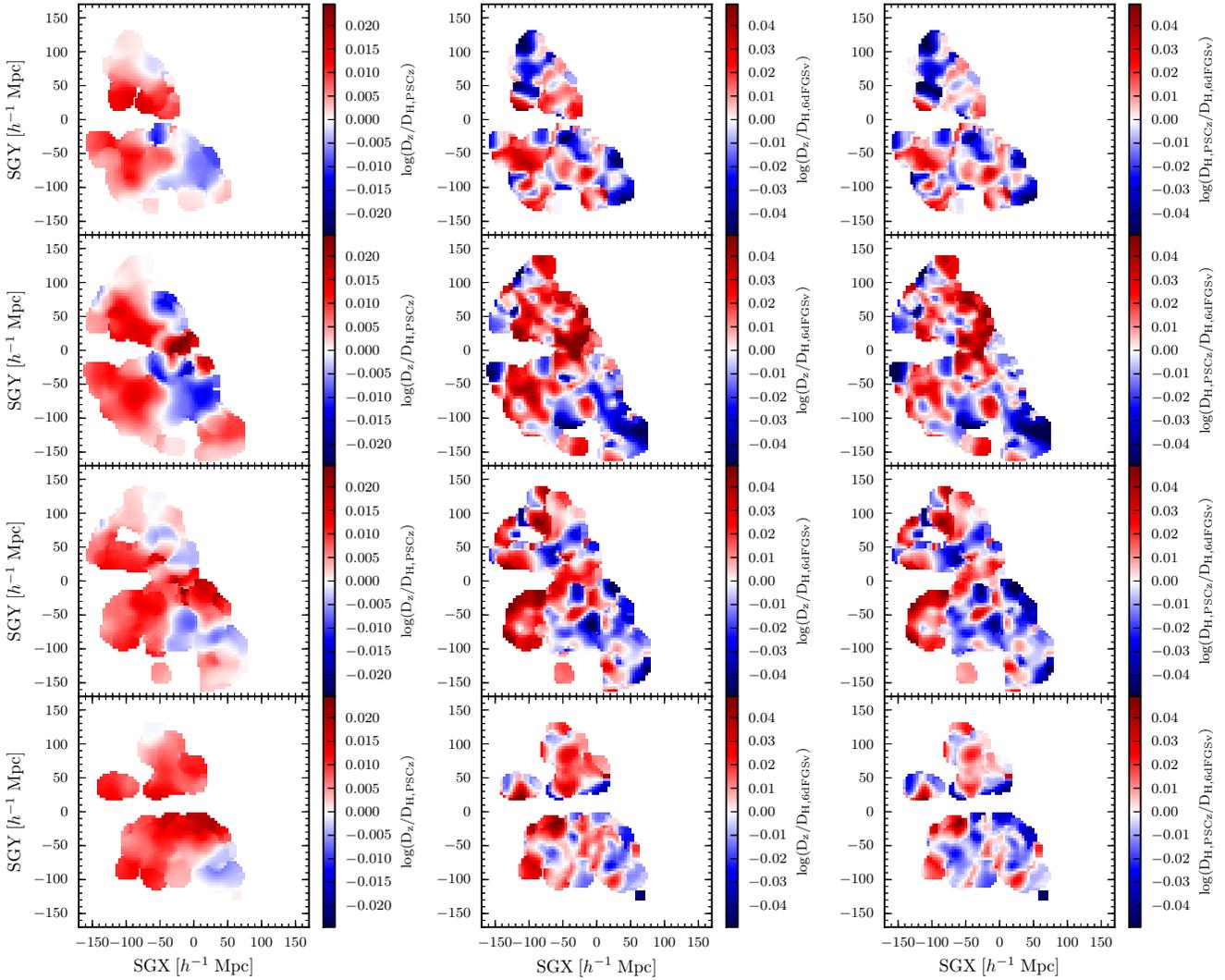} 
\caption{ Same as Figure 9, but with the PSCz velocity field (\citealt{branchini99}) in place of the  2MRS field.\label{FIG10}} 
\end{minipage} 
\end{figure*}

In addition to displaying the velocity fields in SGZ slices as in Figures 9 and 10, we would also like to view the
fields in a fully 3D manner. Figure~11\footnote{In the 3D version of this paper, this plot is an
  interactive 3D visualization, generated using custom C~code and the
  S2PLOT graphics library \citep{barnes06x} following the approach
  described in \citet{barnes08x}. View and interact with this 3D figure
  using Adobe Reader Version 8.0 or higher.} shows the smoothed 3D
6dFGSv peculiar velocity field.

We note that because of the adaptive smoothing, the mean error on the $\Delta d$ value for a given gridpoint is relatively uniform across the survey volume.  We find that the mean error, averaged over all gridpoints, is 0.02 dex in the 3D grid.  However, because Figures 9 and 10 involve additional averaging of gridpoints, in that we collapse the grid onto four SGZ slices, the mean $\Delta d$ error in those plots is 0.009.  Thus features in that plot that vary by less than $\sim 0.009$ may simply be products of measurement uncertainties.

\begin{figure*} \centering 
\begin{minipage}{175mm}
\includegraphics[width=0.75\textwidth]{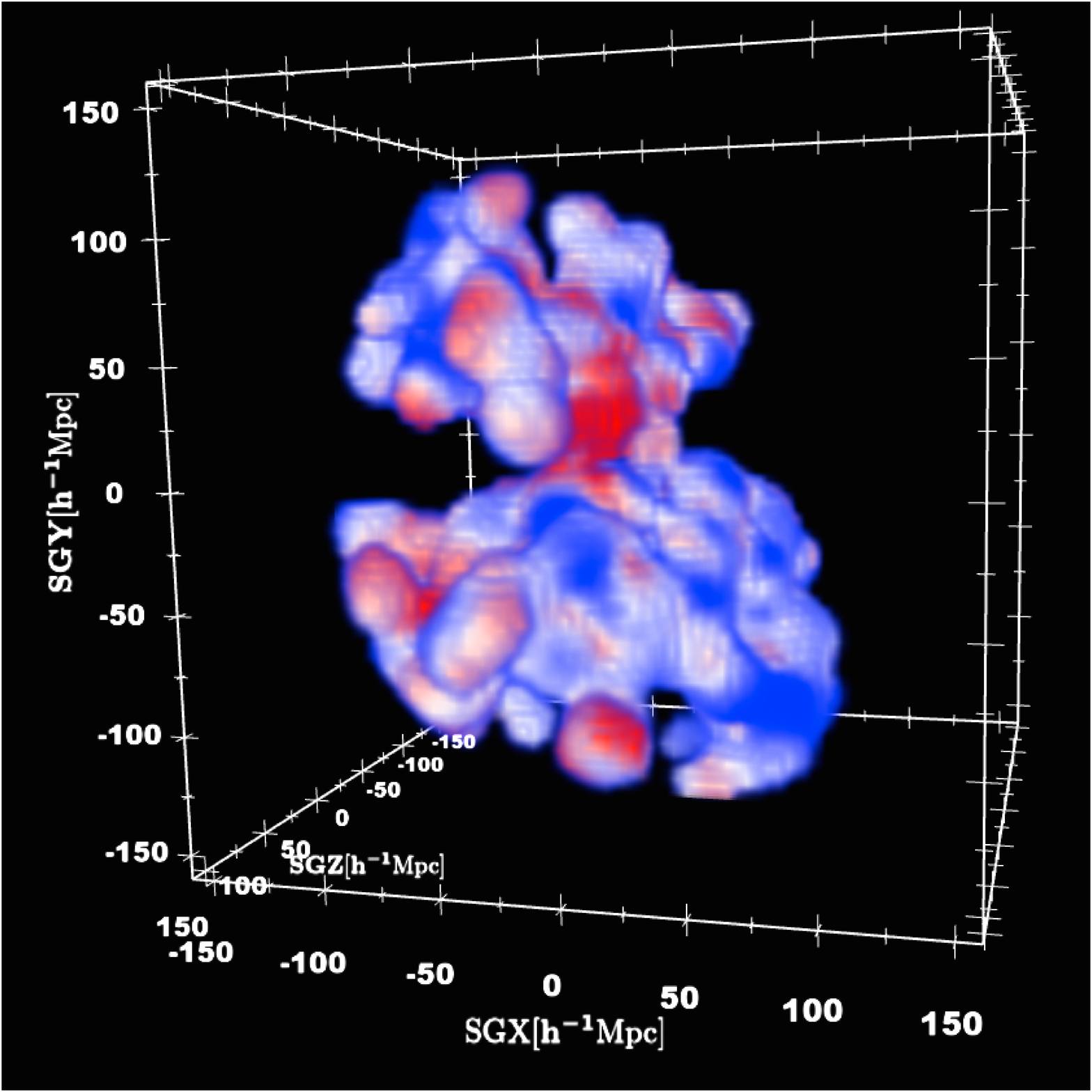}
\hspace{-3.0cm}
\includegraphics[width=0.55\textwidth]{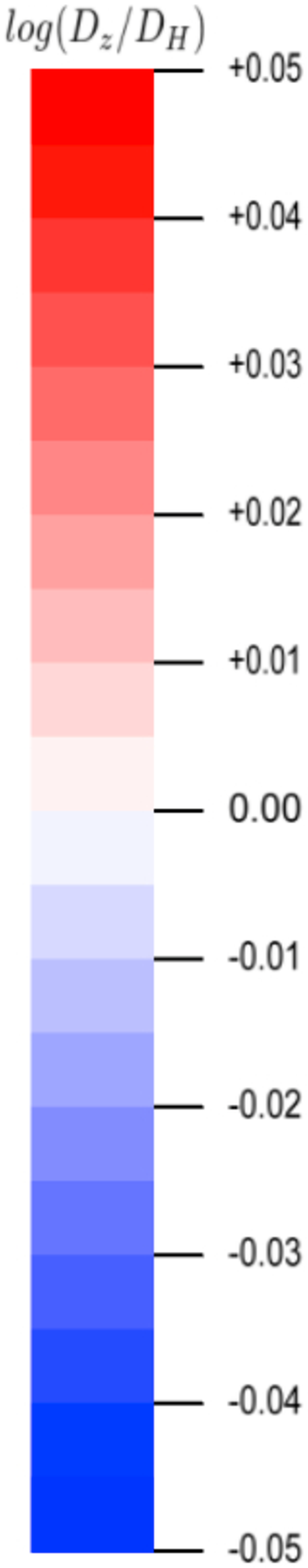}
\caption{The smoothed 6dFGSv peculiar velocity field in 3D, plotted on a grid in supergalactic cartesian coordinates, with gridpoints color-coded by the value of $\Delta d=\log(D_z / D_H)$. In the 3D version of this paper, Adobe Reader version 8.0 or higher enables interactive 3D views of the plot, allowing rotation and zoom.
\label{FIG11}} 
\end{minipage} 
\end{figure*}

\subsubsection{Velocity field `monopole'}

One must be careful in defining the terminology of the velocity moments when considering an asymmetric survey volume, such as the hemispheric volume observed by 6dFGS.  In general, the zeroth order moment of the velocity field, or `monopole', cannot be measured by galaxy peculiar velocity surveys.  This is because the calibration of the velocity field usually involves an assumption about the zeropoint of the distance indicator which is degenerate with a monopole term.  The same logic applies to velocity field reconstructions, such as the 2MRS and PSCz reconstructions used in this paper.

In Section 2.2, we noted that the mean peculiar velocity of gridpoints in the 2MRS reconstruction is +66 \kms .  This value is of course dependent on the fact that we have assumed that the average gravitational potential is zero along the surface of a sphere of radius 200 \mpc .  We now note that for the particular set of gridpoints located at the redshift space positions of galaxies in our sample, the mean is actually somewhat more positive: +161 \kms , with an rms of 297 \kms.  When converted into the logarithmic units of $\Delta d$ and smoothed onto the 3D grid shown in Figure 9, we find a mean value of $\langle\Delta d\rangle=+0.007$ dex for the smoothed 2MRS gridpoints.  This is close to the mean value of $\langle\Delta d\rangle=+0.005$ found in the smoothed 6dFGS gridpoints.  Similarly, for PSCz, the mean peculiar velocity of all gridpoints is +79 \kms, while the mean at the positions of our 6dFGS galaxies is +135 \kms, with an rms of 172 \kms.  This corresponds to $\langle\Delta d\rangle=+0.005$.  That is, in both the 2MRS and PSCz predictions {\it and} in the 6dFGS observations, we find that the mean peculiar velocities at the redshift space positions of the galaxies in our sample skew somewhat towards positive values.

This is {\it not}, however, indicative of a monopole in the velocity field, as our survey only covers the southern hemisphere.  Rather, it is an indication that the model predicts a net positive mean motion of galaxies in the southern hemisphere, at least within the hemisphere of radius $\sim 160$ \mpc \ covered by the survey, and that our observations show a similarly positive mean motion of galaxies in the same hemispheric volume.  (And of course, the latter result depends on the assumption that the mean logarithmic comoving distance ratio, $\langle\Delta d\rangle$, is zero along a great circle in the celestial equatorical region.)

While the mean value of $\langle\Delta d\rangle$ is the same for both the predicted and observed fields, the standard deviation is not.  As noted in Section 4.3, the scatter in $\langle\Delta d\rangle$ for the 6dFGSv galaxies is 0.112 dex.  With the adaptive kernel smoothing, this scatter is reduced to 0.023 dex, whereas for the smoothed 2MRS and PSCz predicted fields, the scatter is only 0.009 and 0.007 dex respectively.  So, while the three fields have the same mean value for $\langle\Delta d\rangle$, the $\langle\Delta d\rangle$ values in the predicted field have a scatter which is comparable to their mean offset from zero, resulting in very few points with negative values.  The scatter is much larger in the observed field, resulting in many more gridpoints with negative values.

The offset of $\langle\Delta d\rangle$ from zero then, does not necessarily indicate the existence of a velocity field monopole, but may simply reflect the existence of higher order moments such as the dipole, with net positive motion towards the southern hemisphere.  We consider the velocity field dipole in the context of the origin of the bulk flow in the following subsection.

\subsubsection{Velocity field dipole and comparison with models}

Measurements of the peculiar velocity field dipole, or `bulk flow', have been a source of some controversy in recent years.  Despite differences in the size and sky distribution of the various peculiar velocity catalogs, there is general agreement among authors on the {\it direction} of the bulk flow in the local universe.  For example, \citet{watkins09}, \citet{nusser11}, and \citet{turnbull12}, among others, all find a bulk flow whose direction, in supergalactic coordinates, points towards $sgl \sim 160$ degrees, $sgb \sim -30$ degrees, roughly between the Shapley Supercluster and the Zone of Avoidance.

Disagreement remains, however, about the {\it magnitude} of the bulk flow, and the extent to which the value may be so large as to represent a disagreement with the standard model of $\Lambda$CDM cosmology.  \citet{watkins09}, for example, claim a bulk flow of $\sim 400$ \kms on a scale of 50 \mpc, which is larger than predicted by the standard $\Lambda$CDM parameters of WMAP \citep{hinshaw13} and Planck \citep{planck13}.  Others, such as \citet{nusser11}, claim a smaller value that is not in conflict with the standard model.

If the bulk flow is larger than the standard cosmology predicts, then it may be because the standard cosmological picture is incomplete.  In a `tilted universe' \citep{turner91}, for example, some fraction of the CMB dipole is due to fluctuations from the pre-inflationary universe.  In that case, we would expect to observe a bulk flow that extends to arbitrarily large distances. (Though it should be noted that the results of \citealt{planck13b} cast the plausibility of the titled universe scenario into doubt.)

However, a large bulk flow could instead have a `cosmographic' rather than a `cosmological' explanation.  The geometry of large scale structure near the Local Group may be such that it induces a bulk flow that is much larger than would typically be seen by a randomly located observer.  In particular, there has been debate regarding the mass overdensity represented by the Shapley Supercluster (e.g., \citealt{hudson03}, \citealt{proust06}, \citealt{lavaux11}), which, as seen in Figure 3, represents the most massive structure within $\sim$ 150 \mpc.  (We should note, however, that the dichotomy between a cosmological and cosmographic explanation for a large bulk flow expressed above is somewhat incomplete.  A cosmographic explanation could have its own cosmological origins, in that a deviation from $\Lambda$CDM could impact the local cosmography.  Nonetheless, certain cosmological origins for the bulk flow, such as a tilted universe, would not necessarily have such an impact on the cosmography.)

Whether we are able to identify the particular structures responsible for the bulk flow thus bears on what the origin of the large bulk flow might be.  Most previous datasets were shallower than 6dFGSv, so this is of particular interest in this case.  Our survey volume covers most of the Shapley Supercluster, allowing us to compare the predicted and observed velocities in the Shapley region.  In Magoulas et al. (in prep) and Scrimgeour et al. (submitted), we will make quantitative measurements of both the bulk flow and the `residual bulk flow' (the component of the velocity dipole not predicted by the model velocity field), but those results will be informed by our cosmographic comparison here.

The first question is whether the velocity field models provide a good fit to the velocity field data.  For each of the 6dFGSv galaxies, we fit the $\Delta d$ probability distributions to a Gaussian function.  We define $\Delta d^{data}$ as the mean value of $\Delta d$ in the Gaussian fit, and $\epsilon$ as the width of the Gaussian.  The corresponding $\Delta d$ from either the 2MRS or PSCz models is then $\Delta d^{model}$.  We then define the reduced $\chi^2$ statistic, 
\be
\chi_{\nu}^2 =  \sum_{n=1}^N \left[{\Delta d^{data}_n-\Delta d^{model}_n \over \epsilon_n}\right]^2 / N
\ee
for the $N=8885$ galaxies in the sample.  We find $\chi^2_{\nu}=0.897$ for 2MRS and $\chi^2_{\nu}=0.893$ for PSCz.  Both values are $\sim1$, and thus represent a good fit of the data to the model.  This is not surprising---because the uncertainties in the observed peculiar velocities are substantially larger than the predicted velocities, comparisons are bound to yield $\chi^2_{\nu} \sim 1$.  However, we note that the FP scatter as measured in \citet{magoulas12x} assumes that the 6dFGS galaxies are at rest in the CMB frame, and so $\chi^2_{\nu} \sim 1$ by construction.  The fact that both 2MRS and PSCz show smaller values of $\chi^2_{\nu}$ thus indicates an improvement over a model in which the galaxies have no peculiar velocities at all.  To compare between the models, we look at the total $\chi^2$, $\chi^2_{tot}=N\chi^2_{\nu}$.  In this case, $\chi^2_{tot}=7970$ for 2MRS and $\chi^2_{tot}=7934$ for PSCz.  PSCz is thus the preferred model with high significance.

Rather than simply compute a global $\chi^2_{\nu}$, we can also investigate the agreement between data and model along particular lines of sight.  Note that most of the Southern Hemisphere structures highlighted in Figure 3 lie roughly along two lines of sight, $\sim 130$ degrees apart.  Hereafter, we refer to these directions as the `Shapley direction' (the conical volume within 30 degrees of $(sgl,sgb)=(150.0^\circ ,-3.8^\circ$) ) and the `Cetus direction' (the conical volume within 30 degrees of $(sgl,sgb)=(286.0^\circ ,+15.4^\circ$)).  These sky directions correspond to the positions of the more distant concentration of the Shapley Supercluster and the Cetus Supercluster, as identified in Figure 3, respectively.

We can see the velocity flows along both of these directions in 3D in Figure 11.  However,  even with such an interactive plot, one cannot easily see deep into the interior of the survey volume.  To mitigate this problem, we have created Figure 12, which is identical to Figure 11, except that only certain gridpoints are highlighted.  In this figure, we display only those gridpoints with extreme values of $\Delta d$ (greater than +0.03 or less than -0.03 dex).  We also highlight the positions of each of the superclusters highlighted in Figure 3, in addition to the position of the Vela Supercluster (see below).

\begin{figure*} \centering 
\begin{minipage}{175mm}
\includegraphics[width=0.75\textwidth]{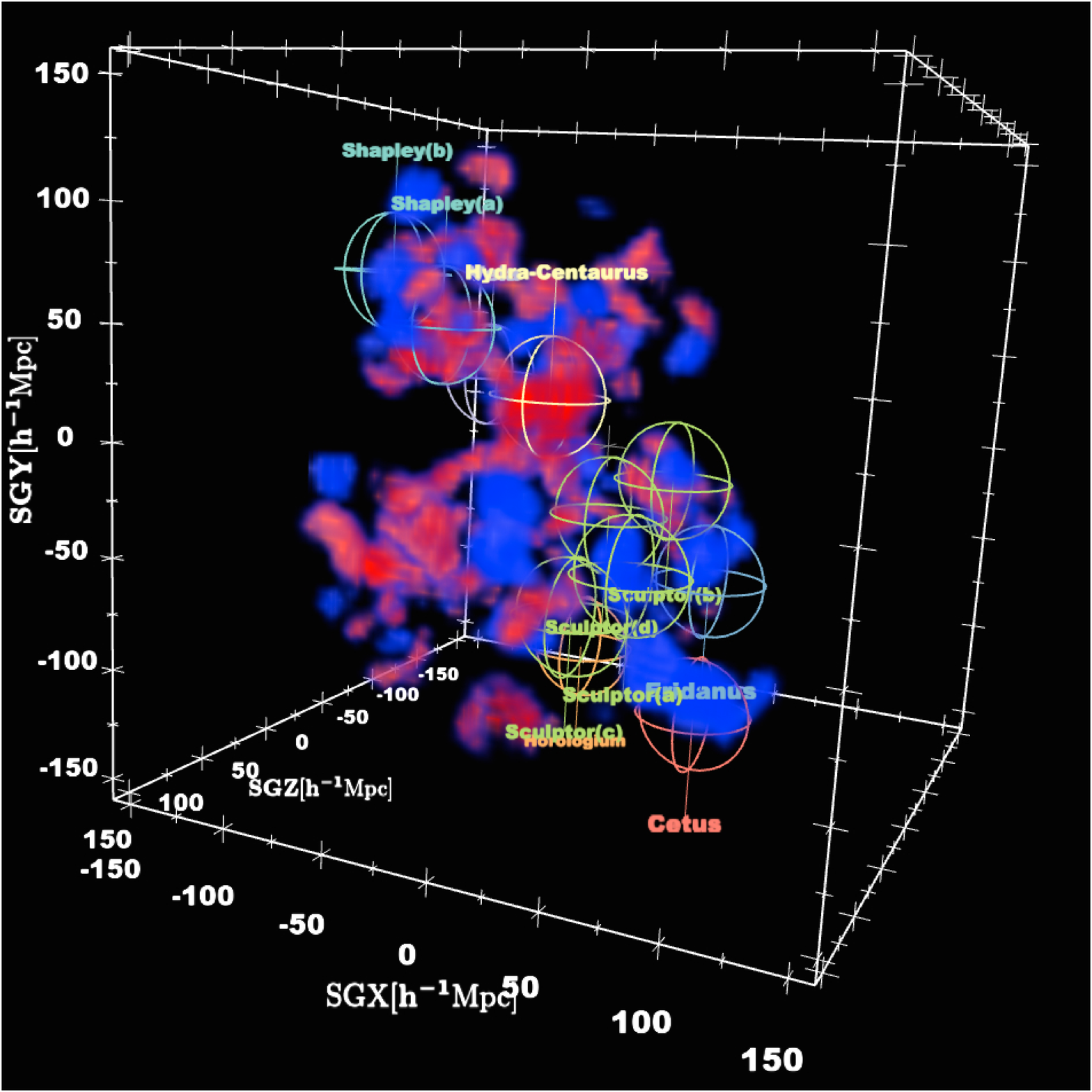}
\hspace{-3.0cm}
\includegraphics[width=0.55\textwidth]{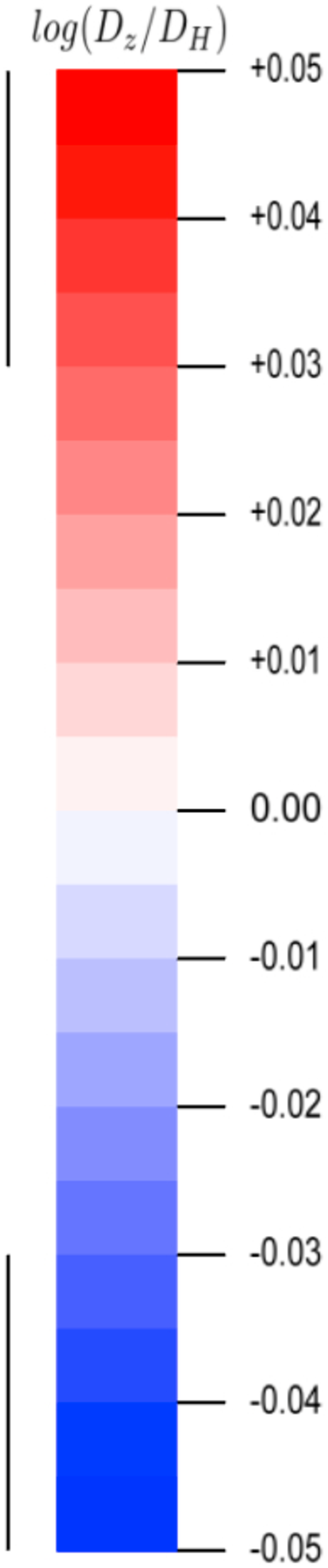}
\caption{Same as Figure 11, except that we only show gridpoints with 
$\Delta d=\log(D_z / D_H)$ either greater than +0.03 or less than -0.03, in order to highlight the regions with the most extreme values.  We also label each of the features of large scale structure labeled in Figure 3.  Though they are outside the survey volume, we include labels for the Vela and Horologium-Reticulum superclusters, as they exert influence on the local velocity field.  In the 3D version of this paper: To toggle
the visibilty of individual superclusters in the interactive figure (Adobe Reader), open the Model Tree, expand the
root model, and select the required supercluster name.
\label{FIG12}} 
\end{minipage} 
\end{figure*}

As seen in these figures, we find mostly positive peculiar velocities along the Shapley direction, and negative peculiar velocities along the Cetus direction.  Does this agree with the models?  In the Shapley direction alone, $\chi^2_{\nu}=0.920$ for 2MRS and 0.917 for PSCz.  Whereas in the Cetus direction alone, $\chi^2_{\nu}=0.914$ for 2MRS and 0.898 for PSCz.  Thus the agreement between data and models is somewhat worse along each of these lines of sight than it is in the survey volume as a whole.

We investigate the agreement between the observations and models further in Figure 13.  As shown in this figure, we have binned the 6dFGSv galaxies in 10 \mpc \ width bins along various directions, including the Shapley and Cetus directions.  In each bin, we average the values of $\Delta d$ for all galaxies in the bin.  We then assign errorbars according to $\epsilon_{bin}$ as a function of the $\epsilon_n$ values of the galaxies within each redshift bin (where $\epsilon_n$ is the same $\epsilon_n$ used in Equation 26), according to:
\be
\epsilon_{bin}^2=\sum_{n=1}^{N_{bin}} \epsilon_n^2 / N_{bin}
\ee
where $N_{bin}$ is the number of galaxies in the bin.  We compare these to the corresponding average values of $\Delta d$ at the redshift space positions of the same galaxies in both the 2MRS and PSCz models.  As we see in this figure, there is a systematic offset in $\Delta d$ in the Cetus direction (a more significant disagreement for 2MRS than for PSCz), with $\Delta d$ being, on average, 0.020 and 0.010 dex lower than the 2MRS and PSCz predictions respectively.  We note that there is a somewhat smaller systematic offset in the Shapley direction as well, with $\Delta d$ being, on average, 0.007 and 0.005 dex higher than the 2MRS and PSCz predictions respectively.

\begin{figure*}
\begin{minipage}{175mm}
\includegraphics[width=1.0\textwidth]{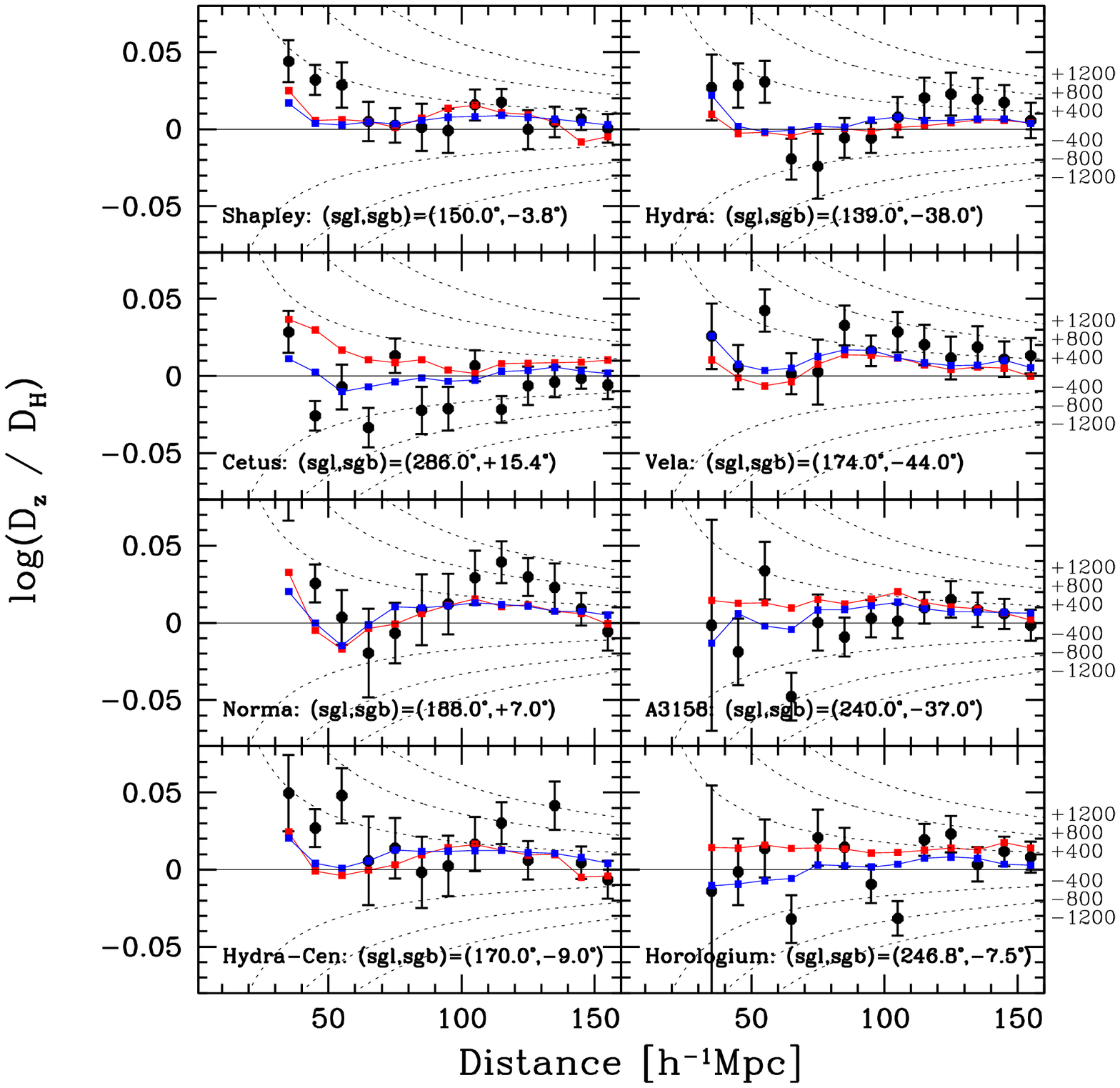}
\caption{Averaged $\Delta d=\log(D_z / D_H)$ for 6dFGS observations, as compared to both the 2MRS and PSCz models, in redshift space distance bins along the Shapley Supercluster, Cetus Supercluster, Norma Cluster, Hydra-Centaurus Supercluster, Hydra Cluster, Vela Supercluster, Abell 3158, and Horologium-Reticulum Supercluster directions.  Each bin is 10 \mpc \ wide, and the directions are defined as the regions within 30 degrees of the coordinates listed at the bottom of each panel.  For each bin, we have averaged the values of $\Delta d$ for all 6dFGSv galaxies within the bin, and display the averaged value as the black circle.  The errorbar is then given by Equation 27.  The averaged $\Delta d$ values as given by the 2MRS and PSCz models are represented by the red and blue squares respectively.  Red and blue lines connect the points.  We also draw a black line at $\Delta d = 0$, and dotted lines to show the $\Delta d$ values corresponding to $+/- 400$, 800, and 1200 \kms, as indicated along the righthand side of the plot.
\label{FIG13}}
\end{minipage}
\end{figure*}

As a point of comparison, we have generated similar plots for several additional lines of sight, shown in the remaining panels in Figure 13.  We show the binned $\Delta d$ values along the directions of the Norma Cluster, the Hydra-Centaurus Supercluster, the Hydra Cluster, and the Vela Supercluster.  The first three structures are familiar features of the local large scale structure, noted by numerous past authors (e.g., \citealt{lynden-bell88}, \citealt{tully92}, \citealt{mutabazi14}).  Vela is less well know, but Kraan-Korteweg et al. (in prep, private communication) find preliminary observational evidence for a massive overdensity in that direction at $cz \sim 18,000-20,000$ \kms.  Each of these four sky directions lies closer to the Shapley direction than the Cetus direction.  They also lie close to both the Zone of Avoidance and the bulk flow directions observed by various authors, such as \citet{feldman10}, \citet{nusser11}, and \citet{turnbull12}.  Additionally, they each show a similar trend to the one seen in the Shapley direction: The $\Delta d$ values lie above the model predictions from both 2MRS and PSCz.



The remaining two panels in Figure 13 show the velocity field along the directions towards Abell 3158 and the Horologium-Reticulum Supercluster.  These are much closer to the Cetus direction than the Shapley direction, and they show a similar trend to the one seen for Cetus: $\Delta d$ values which lie {\it below} the model predictions from both 2MRS and PSCz.  Like Cetus, they also show a somewhat larger divergence between the 2MRS and PSCz model predictions, with PSCz lying closer to our observed $\Delta d$ values.


These plots confirm what can be seen in  Figures 9 and 10 as well.  There is a gradient of residuals from the model, going from somewhat negative residuals in the Cetus direction, to more positive residuals in the Shapley direction, with the Cetus direction representing a particularly large deviation between the data and model for 2MRS, at least in terms of the mean value of $\Delta d$, even if the $\chi^2_{\nu}$ value in that region is no worse than the corresponding value in the Shapley direction..  This suggests a residual bulk flow from both the 2MRS and PSCz models, pointing in the vicinity of the Shapley Supercluster, which is explored in greater detail by Magoulas et al. (in prep).


One might worry that the apparent direction of this residual bulk flow lies close to the Galactic plane.  Might erroneous extinction corrections be creating a systematic bias, which skews our results?  As noted in Section 2.1, a previous iteration of this catalog made use of the \citet{schlegel98} extinction map rather than the \citet{schlafly11} extinction map.  We find virtually no change in the cosmography, when using the \citet{schlegel98} corrections rather than those of \citet{schlafly11}.  Magoulas et al. (in prep.) investigates this issue further, measuring the bulk flow when the extinction corrections are changed by as much as three times the difference between the \citet{schlegel98} and  \citet{schlafly11} corrections, finding only small changes in both the magnitude and direction of the bulk flow  between one extreme and the other.  It thus seems unlikely that the apparent residual bulk flow is an artifact of erroneous extinction corrections, unless the \citet{schlafly11} extinction map includes systematic errors across the sky far larger than the difference between those values and those of \citet{schlegel98}.

So, in summary, the global $\chi^2_{\nu}$ is $\sim 1$ for both models, though lower for PSCz than for 2MRS.  Both models systematically predict peculiar velocities that are too negative in the Shapley direction, and too positive in the Cetus direction.  This suggests a `residual bulk flow' that is not predicted by the models.

The residual bulk flow suggests that the models may either be underestimating features of large scale structure within the survey volumes of both 2MRS and PSCz, or that the velocity field is being influenced by structures {\it outside} the survey volume.  However, the fact that the larger discrepancy in mean  $\Delta d$ occurs in the Cetus direction rather than the Shapley direction would seem to argue against a more-massive-than-expected Shapley Supercluster being the main cause of the residual bulk flow, {\it unless} we are underestimating the zeropoint of the FP by much more than the 0.003 dex uncertainty derived in Section 3.2.

We did note in Section 3.2, however, that the calibration of the FP zeropoint depends on the assumption that the sample of galaxies within the equatorial region (those galaxies in the range $-20^{\circ} < decl. < 0^{\circ}$) exhibit no monopole feature in the velocity field.  We could ask, at this point, whether shifting the zeropoint might change the agreement between data and model.

We thus calculate $\chi^2_{tot}$ for both the 2MRS and PSCz models, allowing the zeropoint of $\Delta d$ to vary as a free parameter.  As seen in Figure 14, the best fit zeropoint for 2MRS and PSCz respectively would be +0.0080 and +0.0102 dex higher than our nominal value.  This ``higher zeropoint'' means that the $\Delta d$ values would be correspondingly {\it lower} (i.e., the redshift distance is less than the Hubble distance).  Thus, allowing the zeropoint to float as a free parameter in the comparison to both 2MRS and PSCz models would have the effect of improving the overall fit, but making the offset in the mean value of $\Delta d$ between data and models in the Cetus direction {\it worse}.

\begin{figure}
\includegraphics[width=0.45\textwidth]{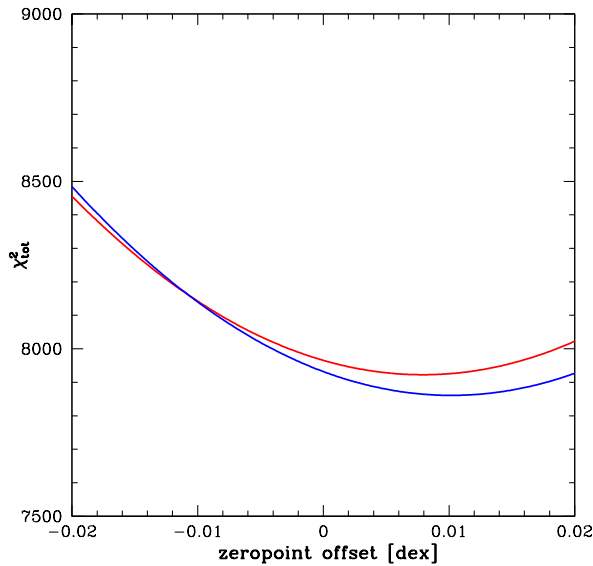}
\caption{$\chi_{tot}^2$ as a function of the zeropoint offset for 2MRS (red) and PSCz (blue).  The best fit zeropoint offsets are +0.0080 and +0.0102 dex for 2MRS and PSCz respectively.  These means that the values of $\Delta d$ would be shifted correspondingly {\it lower} in each case.
\label{FIG14}} 
\end{figure}

Finally, is there anything that we can say about the differences between the two velocity field models, and why PSCz offers a better fit to the 6dFGSv velocities than 2MRS does?  To see why the 2MRS and PSCz velocity fields differ, it is instructive to look at the respective density fields, as shown in Figure 3.

As seen in that figure, while the same basic features of large scale structure appear in both models, they differ in the details, with a mean rms of the log density ratio on a gridpoint-by-gridpoint basis being 0.73 dex.  (The scatter appears somewhat smaller than this in Figure 3, because we have averaged gridpoints at a given SGX, SGY position onto our four SGZ slices.)  The deviations are greatest at the edges of the survey volume, though relatively evenly spread across the sky, with no one particular feature of large scale structure dominating the differences between the models.  Within 161 \mpc, the mean overdensity $\langle\delta\rangle$ is -0.07 in 2MRS and -0.15 in PSCz.  With PSCz being, on average, less dense than 2MRS near the limits of the 6dFGS survey volume, it features more negative peculiar velocities in both the Shapley and Cetus directions, perhaps accounting for some of the better agreement with 6dFGSv in the Cetus direction.

We should note that, as seen in the original 2MRS and PSCz papers (\citealt{erdogdu06}, \citealt{branchini99}), both surveys have very few galaxies at redshifts of $cz\sim 15,000$ \kms and greater, leading to considerable uncertainty in the density/velocity model at those redshifts.  A future paper will improve on this limitation by comparing the observed velocity field to the deeper 2M++ reconstruction \citep{lavaux11}.  In the future, deeper all-sky redshift surveys, such as WALLABY \citep{duffy12} and TAIPAN (\citealt{beutler11}, \citealt{colless13}) should be able to provide more accurate models of both the density and velocity fields at the distance of structures such as Shapley.  Those same surveys will also provide significantly more peculiar velocities than are presently available, which may be enough to resolve the source of any residual discrepancies between data and models.

\section{Conclusions}

We have derived peculiar velocity probability distributions for 8885
galaxies from the peculiar velocity subsample of the 6dF Galaxy Survey
(also known as 6dFGSv). We have presented a Bayesian method for deriving
the probability distributions, which are nearly Gaussian with logarithmic distance. The Bayesian
approach allows us to take advantage of the full probability
distribution, accounting for the fact that it is not perfectly Gaussian
in logarithmic units (and certainly not in linear units).  In the units of the logarithmic distance ratio, $\Delta d$, we find a mean value of $\Delta d$ equal to +0.005, in agreement with the slightly positive values for southern hemisphere galaxies given by the 2MRS and PSCz models.  The mean scatter in $\Delta d$ for individual galaxies is 0.112 dex, corresponding to a 26\% distance error in linear units.

The peculiar velocities are then smoothed using an adaptive Gaussian kernel to give 3D maps of the observed velocity field. We similarly
smooth the 2MRS and PSCz predicted velocity fields, and compare them to the 6dFGSv field.  We find $\chi_{\nu}^2=0.897$ for 2MRS and $\chi_{\nu}^2=0.893$ for PSCz.  The difference in {\it total} $\chi^2$ is 36, favoring the PSCz model with high significance.  Though $\chi_{\nu}^2 \sim 1$ in both cases, the agreement is not uniform across the survey volume.  The observed field shows a stronger dipole signature than is seen in either of the predicted fields, with systematically positive peculiar velocities being found in the vicinity of the Shapley Supercluster, as well as other neighboring structures, such as the Norma Cluster and Vela Supercluster.  Several previous authors (e.g., \citealt{feldman10}, \citealt{nusser11}) have found that the bulk flow of the local universe points in the vicinity of these structures.  We find that these more positive than expected peculiar velocities are offset by more negative than expected peculiar velocities in the direction of the Pisces-Cetus Supercluster (`Cetus direction'), $\sim 130^\circ$ away.

The larger than expected dipole signature across the sky may have either a cosmological or cosmographic origin.  The latter interpretation would suggest that the models either overestimate or underestimate features of large scale structure within the survey volume, or that some features of large scale structure {\it outside} the survey volume have a large impact on the velocity field.  We note that the bulk of the 6dFGSv galaxies lie at distances greater than 100 \mpc, whereas the number counts in both the 2MRS and PSCz surveys peak at nearer distances.  Thus, the contribution to the models from more distant structures is dependent on a comparatively small number of objects.  It does not appear, however, that any mismatch between data and models results from a straightforward underestimate of the Shapley Supercluster in the models, as, though $\chi_{\nu}^2$ in the Shapley direction alone is larger than the global $\chi_{\nu}^2$, the discrepancy in the mean of all logarithmic distances ratios is  greater in the Cetus direction than the Shapley direction.  In fact, when we allow the zeropoint of the FP to float as a free parameter, we find greater agreement between data and models when the observed velocities are pushed towards more negative values, thus making the agreement between data and models in the Cetus direction {\it worse}.

We are currently investigating improved density and velocity field models, to advance our understanding of any discrepancies between data and models.  In the forthcoming Magoulas et al. (in prep), we examine the bulk flow, and residual bulk flow from both 2MRS and PSCz models, in greater quantitative detail.  Additionally, future all-sky redshift surveys will improve the knowledge of the density to a greater depth than can be studied by the current generation of surveys.

\section*{Acknowledgements}

We acknowledge the efforts of the staff of the Australian Astronomical
Observatory (AAO), who have undertaken the survey observations and
developed the 6dF instrument. We thank Alex Merson for providing the
6dFGS group catalogue, Enzo Branchini for providing a copy of the PSCz reconstruction, and Chris Blake and Yin-Zhe Ma for useful comments and discussion.  Three-dimensional visualisation was conducted
with the S2PLOT progamming library \citep{barnes06x}.  D.H. Jones acknowledges support from the Australian Research Council Discover-Projects Grant (DP-0208876), administered by the Australian National University.  J. R. Mould acknowledges support from
the Australian Research Council Discovery-Projects Grant (DP-1092666).  C.~Magoulas has
been supported by a scholarship from the AAO.  J.R. Lucey acknowledges support from the U.K. Science and
Technology Facility Council (STFC, ST/I001573/I).  This research was conducted by the Australian Research Council Centre of Excellence for All-sky Astrophysics (CAASTRO), through project number CE110001020.

\begin{appendix}

\section{The lognormal distribution of peculiar velocities from a
  Gaussian distribution of Fundamental Plane offsets} 

In this appendix, we address the distribution of peculiar velocities
arising from Fundamental Plane distance estimates. Specifically, we
derive the lognormal distribution of peculiar velocities that results
from a Gaussian error distribution for the offsets in logarithmic
distance ratio from the Fundamental Plane. We compare these results with
the original work of \citet{lynden-bell88}, noting similarities and
differences. We discuss the particular biases that arise from the
asymmetry of the lognormal distribution, but leave it up to the reader
to decide how best to account for this in using the 6dFGS dataset for a
particular application.

For simplicity, the following derivation considers the peculiar velocity
distribution of galaxies that derives from their nominal offset from the
Fundamental Plane and its Gaussian uncertainty. It ignores the fact that
for our 3D Gaussian model the maximum likelihood offset for a fixed $i$
and $s$ is not the offset from the Fundamental Plane itself (a matter
discussed in Section~8.3 of \citet{magoulas12x}), as that does not
affect the general argument made here. It also ignores complicating
effects due to selection bias. All these complications are dealt with in
the detailed algorithm used to derive the posterior velocity
distributions discussed in the main text; the point of this appendix is
to derive a simple but relevant analytic result to inform the reader's
understanding.

\subsection{Peculiar velocities from Fundamental Plane offsets}

First we derive from basic principles the relationship between a
galaxy's peculiar velocity and its offset from the Fundamental Plane.

A galaxy's peculiar redshift $z_p$ is related to its observed redshift
$z$ and its Hubble redshift $z_H$ (the redshift corresponding to its
distance) by 
\be 
(1+z) = (1+z_H)(1+z_p) ~. 
\ee
We measure distances from the standard ruler provided by the Fundamental
Plane through the relation
\be
R_\theta = \frac{R_z}{d_A(z)} = \frac{R_H}{d_A(z_H)}
\ee
where $R_\theta$ is the angular size of the galaxy, $R_z$ and $R_H$ are
the corresponding physical sizes if the galaxy is at angular diameter
distances $d_A(z)$ and $d_A(z_H)$ given by the observed and Hubble
redshifts ($R_H$ is the galaxy's true physical size because $z_H$
corresponds to its true distance). In practice we infer $R_z$ from the 
observed redshift as $R_\theta d_A(z)$.

The ratio of the true and observed physical sizes is thus
\be
\frac{R_H}{R_z} = \frac{d_A(z_H)}{d_A(z)} 
= \frac{d(z_H)}{d(z)}\frac{1+z}{1+z_H} = \frac{d(z_H)}{d(z)} (1+z_p)
\ee
where $d(z)$ and $d(z_H)$ are the comoving distances corresponding to $z$
and $z_H$, and we have used the general relations $d_A(z)=d(z)/(1+z)$
and, from Equation~A1, $(1+z_p)=(1+z)/(1+z_H)$.

We infer the (logarithmic) true size from the Fundamental Plane relation
\be
\log R_H = r_H = a(s-\bar{s}) + b(i-\bar{i}) + \bar{r} ~.
\ee
We assume that any offset from the
Fundamental Plane is due to the peculiar velocity, so that
\be
\log R_z = r_z = a(s-\bar{s}) + b(i-\bar{i}) + \bar{r} + \delta ~.
\ee
Thus 
\be
\log R_z - \log R_H = \delta \textrm{~ and so ~} \frac{R_z}{R_H} =
10^\delta ~.
\ee

Up to this point we have made no approximations, but now we make use of
the low-redshift approximation $d(z_H) \approx c z_H/H_0$ (or, more
precisely, the approximation $d(z_H)/d(z) \approx z_H/z$), which turns
Equation~A3 into
\be
\frac{R_H}{R_z} = \frac{d(z_H)}{d(z)} (1+z_p) \approx \frac{z_H}{z} (1+z_p) ~.
\ee
Using Equation~A1 to eliminate $z_H = (z-z_p)/(1+z_p)$ and Equation~A6
for the relation between the distance ratio and the Fundamental Plane
offset we obtain 
\be
\frac{R_H}{R_z} \approx \frac{z-z_p}{z} \approx 10^{-\delta} ~.
\ee
Solving for $z_p$ gives $z_p \approx z(1-10^{-\delta})$, so the inferred
peculiar velocity for a galaxy at observed redshift $z$ having an offset
$\delta$ from the Fundamental Plane is
\be
v_p = cz_p \approx cz(1-10^{-\delta}) ~.
\ee

This is the standard approximate relation for the peculiar velocity
based on the low-redshift Hubble law---see, for example,
\citet{lynden-bell88} and \citet{colless01a}. Note that $\delta$
corresponds to $\langle\Delta d\rangle$, the mean logarithmic distance
ratio given in Table~1 \citep[$\delta$ here has the opposite sign
convention to that adopted in][]{colless01a}.

In determining the 6dFGS peculiar velocities we in fact use the exact
distance relation, but this approximation provides a simple and precise
analytic formula to work with. If $z_H/z=(1+\epsilon)d(z_H)/d(z)$ then
$cz_p = (1+\epsilon)cz(1-10^{-\delta})$, and so the relative error in
the peculiar velocity is $\Delta cz_p/cz_p = \epsilon$. Direct numerical
comparison with the exact relation shows that the approximation is very
good: the resulting relative error in peculiar velocity is less than 5\%
at all redshifts (i.e.\ less than 15\,\kms\ for a peculiar velocity of
300\,\kms\ and less than 50\,\kms\ for a peculiar velocity of
1000\,\kms), and less than 1\% for all $cz>3000$\,\kms.

\subsection{The lognormal distribution of peculiar velocities}

As we have noted both in this paper and in our investigation of the
properties of the Fundamental Plane (Magoulas et al.\ 2012), the error
distributions for the offsets of galaxies from the Fundamental Plane 
(combining observational errors and intrinsic scatter about the
relation) are very closely approximated by a Gaussian. Equations~A8
and~A9 then imply that the posterior distributions of relative distances
and peculiar velocities inferred from the Fundamental Plane offsets will
have lognormal distributions.

To derive the peculiar velocity distribution corresponding to a Gaussian
distribution $\mathrm{N}(\delta|\mu,\sigma)$ for the Fundamental Plane
offset $\delta$, we note that the quantity $u$ given by $u=e^{-\delta}$
is lognormal distributed as $\mathrm{ln[N}(u|\mu,\sigma)]$, with
\be
P(u) = \mathrm{ln[N}(u|\mu,\sigma)] = 
\frac{1}{\sqrt{2\pi}u\sigma} \exp{\frac{-(\ln u - \mu)^2}{2\sigma^2}} ~.
\ee
This means that the peculiar velocity given by
\be
v = cz(1-10^{-\delta}) = cz(1-e^{-\delta\ln10}) = cz(1-u^{\ln10})
\ee
is distributed as  
\be
P(v)=P(u)\left|{\frac{du}{dv}}\right| ~.
\ee
By Equation~A11 we have
\be
u=(1-v/cz)^{\frac{1}{\ln10}} ~,
\ee
and thus
\be
\left|{\frac{du}{dv}}\right| = \frac{1}{cz\ln10} (1-v/cz)^{\frac{1}{\ln10}-1} ~.
\ee
Inserting these expressions for $u$ and $|du/dv|$ into Equation~A12, we
obtain 
\begin{align}
P(v) & = \frac{1}{\sqrt{2\pi}\sigma_v(cz-v)} 
         \exp{\frac{-(\ln(cz-v) - \mu_v)^2}{2\sigma_v^2}} \notag \\
     & = \mathrm{ln[N}(cz-v|\mu_v,\sigma_v)]
\end{align}
where $\mu_v = \ln(cz10^\mu) = \ln(cz)$, in the usual case where the
error distribution has $\mu=0$, and $\sigma_v = \sigma\ln10$. 

Hence the peculiar velocities have a lognormal distribution in $cz-v$,
which is the Hubble approximation to the comoving distance in velocity units
($H_0d(z_H) \approx cz_H \approx cz - cz_p = cz - v$); for $v \ll cz$
this is a good approximation.

The mean of this lognormal distribution is 
\be
\mathrm{Mean}[cz-v] = \exp(\mu_v+\sigma_v^2/2)
                    = cz10^{\frac{1}{2}\sigma^2\ln10} ~,
\ee
implying
\be
\mathrm{Mean}[v] = cz(1-10^{\frac{1}{2}\sigma^2\ln10}) ~.
\ee
The standard deviation is
\begin{align}
\mathrm{SD}[cz-v] & = \mathrm{Mean}[cz-v]\sqrt{\exp(\sigma_v^2)-1} \notag \\
                  & = \mathrm{Mean}[cz-v]\sqrt{10^{\sigma^2\ln10}-1} ~,
\end{align}
implying
\be
\mathrm{SD}[v] = cz10^{\frac{1}{2}\sigma^2\ln10}\sqrt{10^{\sigma^2\ln10}-1} ~.
\ee

From Equation~A17, even if $\mu=0$ the mean peculiar velocity is non-zero
and depends on the scatter about the Fundamental Plane. For example, for
the canonical 20\% scatter about the Fundamental Plane we would have
$\sigma=0.08$\,dex, and in that case $\bar{v}/cz = 1-10^{0.08^2\ln10/2}
\approx -1.7\%$, which corresponds to $-$170\,\kms\ if $cz$=10,000\,\kms.

\citet{lynden-bell88} (hereafter LB88) obtained a similar result when
deriving the radial velocity distribution at a given distance
corresponding to an offset from the $D_n$--$\sigma$ relation (a close
relative of the Fundamental Plane). However the approximation they
provide (\citetalias{lynden-bell88} Equation~2.9) is a Gaussian distribution with mean (\citetalias{lynden-bell88}
Equation~2.11) and standard deviation (\citetalias{lynden-bell88} Equation~2.10) identical to
those given above for the lognormal distribution (allowing for
differences in nomenclature and ignoring complications due to Malmquist
bias and intrinsic scatter about the Hubble flow).

In fact, \citetalias{lynden-bell88} do not appear to have realised that the velocity
distribution is actually lognormal. They certainly do not explicitly
identify it as such, even though they derive the first four moments
(\citetalias{lynden-bell88} Appendix~D). They neglect the distribution's skewness and kurtosis
in adopting a Gaussian approximation, arguing that the deviations from
Gaussian form are not significant. While this may be true at small
distances, the effect becomes significant at the distances of most of
the galaxies in the 6dFGS sample. Moreover, the cumulative effect of the
small asymmetries in the peculiar velocity distributions can have a
significant biasing effect on the likelihood of the sample as a whole,
and must be properly accounted for in a careful analysis of this
dataset.

\end{appendix}

\end{document}